\documentclass[a4paper]{spie}  

\usepackage{multirow}
\usepackage[table,xcdraw]{xcolor}
\usepackage{cancel}
\usepackage{physics}
\usepackage{bm} 
\usepackage{amsmath,amsfonts,amssymb}
\usepackage{graphicx}
\usepackage[colorlinks=true, allcolors=blue]{hyperref}   
\usepackage{ulem}
\usepackage{xcolor}

\title{Charge-Spin Conversion in Two-Subband Quantum Wells with Conventional and Unconventional Rashba Spin-Orbit Coupling}

\author[a]{Gerson J. Ferreira}
\author[b]{Boyu Wang}
\author[b]{Jiyong Fu}
\author[c]{Roberto Raimondi}

\affil[a]{Instituto de Física, Universidade Federal de Uberlândia, Uberlândia, Minas Gerais 38400-902, Brazil}
\affil[b]{Department of Physics, Qufu Normal University, Qufu, Shandong 273165, China}
\affil[c]{Dipartimento di Matematica e Fisica, Roma Tre University, via della Vasca Navale 84, 00144 Rome, Italy}

\authorinfo{Further author information: (Send correspondence to Roberto Raimondi.)\\ Roberto Raimondi.: E-mail: roberto.raimondi@uniroma3.it}

\begin{document} 
\maketitle
\normalem

\begin{abstract}
The reciprocal interconversion between spin polarization and charge current (CSC) is the focus of intensive theoretical and experimental investigation in spintronics research. Its physical origin stems from the Rashba spin-orbit coupling (SOC) induced by the breaking of the structure inversion symmetry. The steady-state interconversion efficiency is the result of the non-trivial spin textures of the electric- field distorted Fermi surface. Its full understanding and evaluation requires the consideration of disorder-induced relaxation effects in the presence of spin-orbit induced band splitting. In this paper the additional effect of the orbital degree of freedom is analyzed in a two-subband quantum well with both conventional and unconventional Rashba SOC in the presence of disorder impurity scattering. The latter is treated at the level of the  Born approximation in the Green's function self-energy and with the inclusion of vertex corrections in the linear response functions for the charge current and the spin polarization. By explicitly considering the symmetry properties of the Hamiltonian the matrix structure of the correlation functions is shown to decompose in independent blocks of symmetry-related physical observables. We find that the inclusion of vertex corrections is important for the correct estimate of the CSC efficiency, which also depends on the position of the Fermi level. We also find that the relative sign of the Rashba SOC in the two subbands plays a key role in determining the behavior of the CSC. Finally, we point out how the two-subband model compares with the standard single-band two-dimensional electron gas.
\end{abstract}


\section{INTRODUCTION}
\label{sec:intro}  

The spin-orbit (SO) interaction couples electron spin and
its momentum, affording  electric control and manipulation of magnetic degrees of freedom, the spin, in quantum spintronics~\cite{awschalom:2002,zutic:2004}.
Also, the SO effects underlie novel topological phenomena in diverse fields of quantum condensed matter such as topological insulators~\cite{bernevig2:2006}, Majorana fermions~\cite{lutchyn:2010,oreg:2010}, van der Waals heterostructurs~\cite{geim:2013,gmitra:2015} and Weyl semimetals~\cite{weng:2015,burkov:2015}.
Recent proposals of persistent skyrmion lattice~\cite{fu:2016}, stretchable spin helix~\cite{dettwiler:2017} 
as well as helix-stretch based orbit (pseudospin) filter~\cite{zhao:2023},  which can be 
realized by fine tuning the SO strengths, also indicate the important role of SO effects in semiconductor nanostructures.
Further, the SO field is the key leading to the charge-to-spin conversion by the direct  
Rashba-Edelstein effect (REE) and spin-to-charge conversion by the inverse Rashba-Edelstein effect (IREE)~\cite{edelstein:1990,aronov:1989,ganichev:2002,ivchenko:1978,vorobev:1979,huang:2016,offidani:2017,lin:2019,ghiasi:2019,benitez:2020,monaco:2021}. 
The reciprocal interconversion between spin polarization and charge current plays a crucial role in modern spintronics, and 
accordingly is the focus of intensive theoretical and experimental investigations for spintronic applications.

In semiconductor nanostructures, the SO effects usually have two dominant
contributions, i.e., the Rashba~\cite{rashba:1984} and Dresselhaus~\cite{dresselhaus:1955} terms, arising from the structural and bulk inversion asymmetries,
respectively. While the Dresselhaus coupling mainly depends on the quantum confinement (e.g., the well width)~\cite{walser:2012_2,dettwiler:2017,fu:2015}, the Rashba coupling can be electrically controlled by using an external bias, thus facilitating coherent spin manipulation~\cite{nitta:1997,studer:2009,sasaki:2014,dettwiler:2017}.
As a consequence, the Rashba effect is often used in proposed spintronic devices, e.g., spin-field~\cite{chuang:2015,koo:2009,datta:1990} and spin-Hall effect~\cite{sinova:2015,wunderlich:2010} transistors as well as
spin-charge conversion based applications. 
Extensive studies have been devoted to coherent spin control by resorting to the Rashba SO coupling in semiconductor heterostructures  with only one occupied electron subband~\cite{dettwiler:2017,calsaverini:2008,fu:2020,hao:2015}. 
However, for the case of the single-subband occupancy, the two lifted spin branches of the energy dispersion feature opposite
chiralities, greatly suppressing the efficiency of charge-spin conversion in spintronic applications.  
Recently, following that additional orbital degrees of freedom  may offer more intriguing possibilities for SO control, e.g., band crossing and anticrossing assisted spin manipulation, the spin features in two-band  Rashba quantum systems have also attracted growing interest, with both intra- and interband SO terms \cite{kunihashi:2016, Ferreira2017RW, dettwiler:2017, weigele:2020, deassis:2021, Sergio2023OrbitalEdelstein}.
In general, the intra- and interband SO terms have the same symmetry, leading to the \emph{conventional} 
    two-band Rashba model\cite{bernardes2006spin, EsmerindoPRL2007, calsaverini:2008, fu:2015}.
    Within the conventional model, it has been shown that coupling between subbands can be used to control the spin lifetime in the persistent spin helix regime \cite{fu:2016, dettwiler:2017, Ferreira2017RW, weigele:2020, deassis:2021}. Experimentally, measurements of spin dynamics in two-subband GaAs quantum wells have shown long and anysotropic spin lifetimes \cite{Hernandez2016CISP, Luengo2017Gate, Hernandez2020Anisotropy}.

Recently, Song \emph{et al.} also proposed an \emph{unconventional} two-band Rashba model in two-dimensional (2D) systems~\cite{Song2021}, where the intra- and interband SO terms have different symmetries, so that the Fermi circles of spin-lifted subbands have chirality with the same sign, thus providing high efficiency of spin to charge conversion. 
However, so far, detailed SO features involving the energy dispersion of the four distinct spin branches and the corresponding spin textures for the conventional and unconventional SO models,   which are essential for spintronic applications~\cite{bentmann:2012,noguchi:2017}, still remain obscure.
Further, the effect of vertex correction in momentum space, which is important for the correct charge-spin conversion (CSC)\cite{edelstein:1990}, particularly near the crossing or avoided crossing points of the two bands, is largerly unexplored.

Here, we explore the SO features of the unconventional Rashba model and make comparison to the
conventional one for a QW with two subbands with both intra- and interband SOC. We demonstrate avoided crossings of the energy dispersion and \emph{intertwined} spin textures, in stark  contrast to the conventional Rashba model.  And, near the avoided crossings in momentum space, there may even exist vanishing SO fields, 
triggered by the interband coupling.  This can be used as an handle to suppress spin-relaxation mechanisms for electrons in a 
controllable manner. Furthermore, we take into account the disorder (e.g., impurity) scattering within the Born approximation in the self-energy of Green's function and include the vertex correction of spin textures in the linear response functions for the charge current and spin polarization, to estimate the CSC effciency. We find that 
two distinct regimes can be identified as a function of the Fermi energy, depending whether one or two subbands are occupied. When both bands are occupied, the vertex corrections are important and to a large extent can be understood in a way similar to the behavior of the single-band case. In the regime with only one band occupied, the relevance of the vertex corrections is controlled by the strength of the disorder, e.g. impurity scattering. Our analysis is based on the standard diagrammatic impurity technique here generalized to the case of two bands. Symmetry considerations are exploited in order to simplify the algebraic structure of the vertex corrections arsing from both orbital and spin degrees of freedom and support the numerical evaluation. Within linear response to a d.c. external electric field we evaluate the Kubo formula for both the induced spin polarization and the electrical current, from whose ratio we measure the CSC efficiency.  One byproduct of the algebraic structure of the vertex corrections equation is the information about the spin relaxation times, which appear to behave differently in the conventional and unconventional models.

This paper is organized as follows. In Sec.~\ref{sec:model}, we first present the 
conventional and unconventional two-band Rashba models, as well as the corresponding 
energy dispersion with avoided crossings and novel spin textures. 
The Green's and response functions, involving the self-energy, the vertex equation, and 
the sysmmetry analysis, are introduced in Sec.~\ref{sec:functions}. We present and
discuss numerical results for CSC conversion in Sec.~\ref{sec:numerics}. 
We summarize our  main finding in Sec.~\ref{sec:summary}

\section{Conventional and unconventional Rashba models}
\label{sec:model}
We consider two-dimensional electron gases (2DES) confined in semiconductor quantum wells of two occupied subbands, 
with parabolic dispersion and both intra- and inter-subband Rasbha SO couplings. 
Following the notation introduced in Ref.~\citeonline{Song2021}, we consider the cases of \textit{conventional} and \textit{unconventional} Rashba SOC. The conventional model follows from the usual two-subband GaAs zincblende quantum wells grown along the $z=[001]$ direction~\cite{calsaverini:2008,fu:2015}, where electrons are confined to the $xy$ plane with $x=[110]$ and $y=[1\bar{1}0]$. Assuming structural inversion asymmetry (SIA), the effective Hamiltonian of the conventional model at the $\Gamma$ point must be invariant under the $C_{2V} = \{C_2(z), M_y\}$ point group. In contrast, the unconventional model occurs in 2D systems that transform under the $C_{3V} = \{C_3(z), M_y\}$ point group at the $\Gamma$ point. 

The derivation of both, conventional $H_C$ and unconventional $H_U$, Hamiltonians is shown in Appendix \ref{app:models}. There, we see that $H_C$ and $H_U$ are strikingly similar, differing only by the intersubband Rashba terms $\eta_C$ and $\eta_U$. Consequently, it is useful to recast both Hamiltonians into a generic form,
\begin{align}
    H =
    \begin{pmatrix}
    \varepsilon_{1} & -i\alpha_{1}k_{-} & 0 & -i\eta^* k_{-}
    \\
    i\alpha_{1}k_{+} & \varepsilon_{1} & i\eta k_{+} & 0
    \\
    0 & -i\eta^* k_{-} & \varepsilon_{2} & -i\alpha_{2}k_{-}
    \\
    i\eta k_{+} & 0 & i\alpha_{2}k_{+} & \varepsilon_{2}
    \end{pmatrix}.
    \label{generic_hamiltonian}
\end{align}
For the conventional case, the intersubband Rasbha SO coupling $\eta \equiv \eta_C$ is real, ensuring that 
the intra- ($-i\alpha_{j}k_{-}$) and intersubband ($-i\eta_Ck_{-}$) terms have the same symmetry. 
In contrast, for the unconventional case, $\eta \equiv -i\eta_U$ is imaginary, indicating that the intra- ($-i\alpha_{j}k_{-}$) and intersubband ($\eta_U k_{-}$) terms have different symmetries in spin space. The subband basis is labeled as $\ket{j,\sigma}$, where $j=\{1,2\}$ refer to the subbands, and $\sigma = \{\uparrow, \downarrow\}$ to the spin along $z$. 
We will use two sets of Pauli matrices   $\lambda_0, \lambda_x, \lambda_y, \lambda_z$ and $\sigma_0, \sigma_x, \sigma_y, \sigma_z$ to represent the matrix structure in  subbands and spin degrees of freedom, respectively.
In both cases of intersubband Rashba SO coupling,  the basis $\ket{j,\sigma}$ is sorted as $\{\ket{1\uparrow},\ket{1\downarrow},\ket{2\uparrow},\ket{2\downarrow}\}$, and for each subband $\varepsilon_j = \varepsilon_{j}^{0}+\frac{\hbar^2}{2m}k^2$, $\varepsilon_{j}^{0}$ are the band edges, the effective mass $m$ is assumed to be the same in both subands, $\alpha_j$ is the intra-subband Rashba SO coupling, $\bm{k}=(k_x, k_y)$ is the in-plane quasi-momentum, and $k_\pm = k_x \pm ik_y$. Unless otherwise specified, we use a set of parameters similar to those in Ref.~\citeonline{Song2021}, i.e., 
    $\varepsilon^0_1 = 0$~meV,
    $\varepsilon^0_2 = 160$~meV,
    $m = 0.365m_0$,
    $|\alpha_1| = |\alpha_2| = 78$~meV nm,
    $|\eta_C| = |\eta_U| = 126$~meV nm
, where $m_0$ is the bare electron mass.

\subsection{Subband anti-crossings and spin textures in k-space}

\begin{figure}[ht!]
    \centering
    \includegraphics[width=\columnwidth]{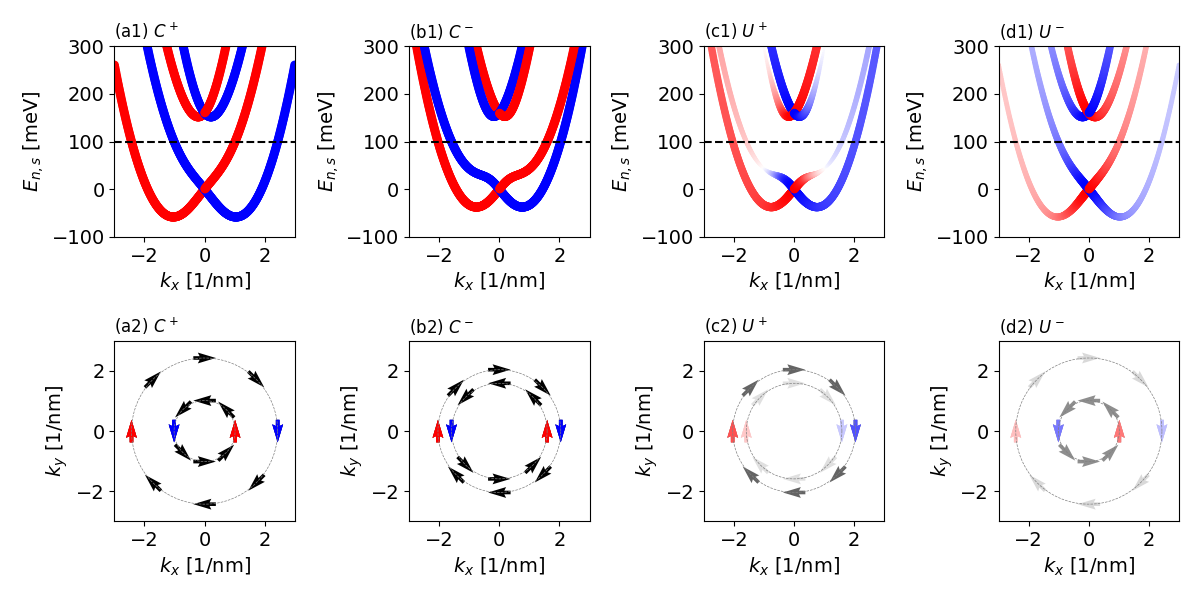}
    \caption{(a1--d1) Band structure for the two-subband models of the conventional ($C^\alpha$) and unconventional ($U^\alpha$) types with $\alpha = \pm$ indicating the sign of $\alpha_1/\alpha_2$ in each panel. The energies $E_{n,s}$ are calculated at $k_y=0$ and the color code refers to the mean value of the spin operator $\sigma_y$ (i.e., $\expval{\sigma_y} > 0$ in blue, and $\expval{\sigma_y} < 0$ in red shades).
    (a2--d2) Spin textures in k-space corresponding to the band structures and energy (dashed line) from the top panels. The arrows and their shade indicate the direction (helicity) and intensity of the spin vector $\expval{\bm{\sigma}}$. At $k_y=0$, the color code matches the one from the top panels.}
    \label{fig:BandSpin}
\end{figure}

The different cases above for $\eta \equiv \eta_C$ or $\eta \equiv -i\eta_U$ yield similar band structures, but they imply different conditions for anti-crossings between the subbands, and distinct spin textures in k-space, as shown in Fig.~\ref{fig:BandSpin}. To see this, first notice that at $k_y = 0$, and with $\eta = 0$, we have $[H,\sigma_y] = 0$. Thus, the wave-functions are eigenstates of $\sigma_y$ with energies $E_{j,\pm} = \varepsilon_j \pm \alpha_j k_x$. If $\alpha_1/\alpha_2 > 0$, the crossing subbands will have the opposite $\sigma_y$, while for $\alpha_1/\alpha_2 < 0$ they have the same $\sigma_y$. Now, for the conventional case, if we consider a finite $\eta \equiv \eta_C$ as a perturbation of the type $H' = \eta_C \lambda_x\otimes\sigma_y \, k_x$, it couples crossing subbands only if they have the same $\sigma_y$ (i.e., $\alpha_1/\alpha_2 < 0$). In contrast, for the unconventional case with $\eta \equiv -i\eta_U$, the perturbation would be $H' = \eta_U \lambda_x \otimes \sigma_x \, k_x$, which couples crossings with opposite $\sigma_y$ (i.e., $\alpha_1/\alpha_2 > 0$). The four possible scenarios lead to the four distinct helical spin textures shown in Fig.~\ref{fig:BandSpin}.

For the conventional cases shown in Fig.~\ref{fig:BandSpin}(a--b), the intersubband SOC $\eta_C$ does not mix the spin components, as argued above, and affects only the band dispersion. In contrast, for the unconventional $\eta_U$ leads to significant spin admixture, as shown by the color code of Fig.~\ref{fig:BandSpin}(c--d). The $U^+$ case, with $\alpha_1/\alpha_2 > 0$, Fig.~\ref{fig:BandSpin}(c) is a particularly interesting scenario, since the spin admixture induced by $\eta_U$ leads to a regime where both spin branches of the lower subband have the same helicity. Consequently, one would expect that this case should lead to a higher charge-spin interconversion, as proposed in Ref.~\citeonline{Song2021}. However, as we will present below, this is not necessarily the case.

\section{The diagrammatic analysis for the Green's and response functions}
\label{sec:functions}
\subsection{Statement of the problem}
\begin{figure}
    \centering
    \includegraphics[width=\columnwidth]{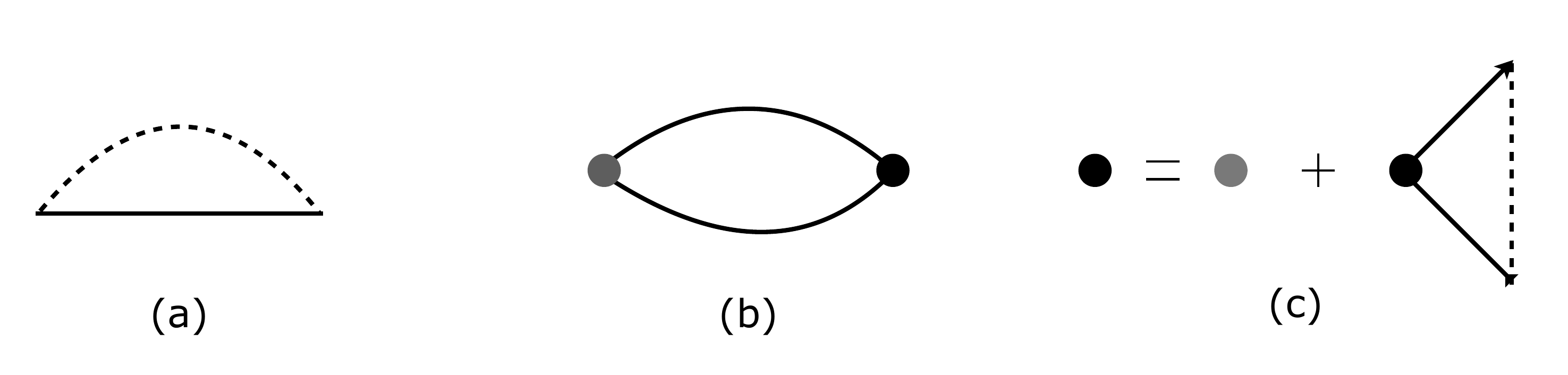}
    \caption{(a) Self-energy {\it rainbow} diagram for the Born approximation. The solid and dashed lines represent the electron Green's function and the impurity average, respectively. (b) {\it Bubble} diagram for the response function. The gray and black dots represent the bare and dressed vertices. (c) {\it Ladder} diagram for the dressed vertex.}
    \label{fig:diagrams}
\end{figure}

Our aim is the evaluation of the  Kubo formula for the response functions of an observable in the presence of an external field. In particular, we will be interested in the spin polarization response to an applied electric field (REE) by allowing electron scattering from random impurities. We will consider standard scalar (i.e. with no dependence on subband and spin degrees of freedom) disorder potential $U({\bf r })$ with delta-like correlated impurities such that $\langle U({\bf r})U({\bf r}')\rangle=u_0^2\delta({\bf r}-{\bf r}')$, where $u_0^2=n_{\rm imp}v_0^2$ with $n_{\rm imp}$ the impurity concentration and $v_0$ the single-impurity scattering amplitude.  At the level of the  Born approximation, the retarded self-energy $\Sigma^R(\omega, {\bf k})$ is 
obtained by the so-called rainbow diagram, as shown in Fig.~\ref{fig:diagrams}(a), and it is
given in terms of the single-particle electron Green's function as
\begin{equation}
    \Sigma^R(\omega, {\bf k})=u_0^2 \int \frac{d^2k'}{(2\pi)^2} G^R(\omega ,{\bf k}')=
    u_0^2 \int \frac{d^2k'}{(2\pi)^2}\sum_{s,n} \frac{P_{s,n}({\bf k}')}{\hbar \omega -E_{s,n} ({\bf k}')+i0^+},
\label{selfenergy}
\end{equation}
where $E_{s,n}({\bf k})$ and $P_{s,n}({\bf k})$ indicate the eigenvalues and the corresponding projector operators of the generic Hamiltonian from Eq.~\eqref{generic_hamiltonian}. The summation run over the possible combinations of the indexes $n=\pm$ and $s=\pm$, which will be defined later on.
Notice that with the adopted model of disorder the self-energy does not depend on the external momentum ${\bf k}$.

Once the self-energy has been evaluated and inserted into the Green's function, the Kubo formula, say, for the y-axis spin polarization in response to an x-axis applied electric field may be evaluated by the diagram shown in Fig.~\ref{fig:diagrams}(b) and reads\cite{edelstein:1990,shen:2014} (below $e>0$ is the unit charge and a factor $\hbar /2$ accounts for the spin value and dimensions)
\begin{equation}
\chi_{yx} =\frac{\hbar}{2\pi}\frac{(-e)\hbar}{2} \int  \frac{d^2k}{(2\pi)^2} {\rm Tr} \left[ S_y^{0} G^R(\omega, {\bf k})J_x ({\bf k})G^A (\omega, {\bf k})\right],
\label{kubo_formula}
\end{equation}
where $S_y^{0}=\lambda_0\otimes \sigma_y$ is the {\it bare } spin-density vertex (in units of $\hbar /2$) and the $J_x ({\bf k})$ is the {\it dressed} charge-current vertex (more precisely, the number-current vertex because we have taken out the charge $-e$). The latter is obtained by considering the ladder diagrams represented in Fig.~\ref{fig:diagrams}(c), which yield the vertex corrections mirroring the self-energy corrections due to the rainbow diagram.
In explicit terms $J_x ({\bf k})$ is obtained by solving the equation\cite{raimondi:2001,schwab:2002,raimondi:2005}
\begin{equation}
J_x ({\bf k}) =J_x^{0} ({\bf k})+u_0^2  \int  \frac{d^2k'}{(2\pi)^2}G^R(\omega ,{\bf k}')
J_x ({\bf k}')G^A (\omega, {\bf k}'),
\label{vertex_equation}
\end{equation}
$J_x^{0} ({\bf k})$ being the {\it bare} charge-current vertex.
The set of 
Eqs.~(\ref{selfenergy},\ref{kubo_formula},\ref{vertex_equation}) are well known\cite{edelstein:1990,raimondi:2001,schwab:2002} and have been applied to the single-subband model with Rashba SOC. Its application to the generic Hamiltonian from Eq.~\eqref{generic_hamiltonian} will be considered in this paper. The presence of two subbands makes the analytic treatment more involved and it is useful to exploit the symmetries of the model to make the solution of the Eqs.~(\ref{selfenergy},\ref{kubo_formula},\ref{vertex_equation}) simpler and physically more transparent. This will be carried out in the following of this section, wheres the discussion of the results will be postponed to a subsequent section.

\subsection{The diagonalization of the Hamiltonian and the structure of the self-energy}
According to Eq.~\eqref{selfenergy}, the evaluation of the self-energy requires the knowledge of the eigenvalues and projection operators of the generic Hamiltonian from Eq.~\eqref{generic_hamiltonian}. We now make the useful observation that the Hamiltonian can be made block-diagonal by introducing the Rashba eigenstates of each subband separately, considering that their spinor structure is independent of the strength of the two Rashba SOC couplings $\alpha_1$ and $\alpha_2$. We then introduce the  new basis by the following unitary transformation
\begin{equation}
S(\theta)=\frac{1}{\sqrt{2}}
\begin{pmatrix}
    1 & -i e^{-i\theta} & 0 & 0\\
     0 &0&   1 & i e^{-i\theta} \\
     1 &i e^{-i\theta} & 0 & 0\\
     0 &0 &1 &-i e^{-i\theta}
\end{pmatrix},
\label{smart_basis}
\end{equation}
where the angle $\theta$ identifies the wave vector momentum ${\bf k}= (k\cos \theta, k\sin \theta)$.
In this new basis, where all quantities will be denoted by an upper index $S$, the Hamiltonian reads
\begin{equation}
H^S \equiv S (\theta) H S^{-1}(\theta)=
\begin{pmatrix}
    \varepsilon_{1}+\alpha_{1}k & i{\rm Im}(\eta) k& 0 &  {\rm Re}(\eta) k \\
  - i{\rm Im}(\eta)  k  & \varepsilon_{2}-\alpha_{2}k & - {\rm Re}(\eta) k & 0\\
  0    &- {\rm Re}(\eta) k  &  \varepsilon_{1}-\alpha_{1}k & -i{\rm Im}(\eta) k \\
 {\rm Re}(\eta) k      &  0 & i{\rm Im}(\eta) k   &  \varepsilon_{2}+\alpha_{2}k
\end{pmatrix}.
\label{smart_hamiltonian}
\end{equation}
The Hamiltonian is block-diagonal if we restrict it to either to the conventional (${\rm Im}(\eta)=0$) or to the unconventional  (${\rm Re}(\eta)=$0) cases. 
Notice as the interband SOC couples equal (opposite) chiralities in the convential (unconventional) case.
We then label the two blocks by
\begin{equation}
h_2^s =
\begin{pmatrix}
\lambda_1^s & \beta_s \\
\beta_s' & \lambda_2^s
\end{pmatrix}, 
\ s=\pm 1,
\label{single_block}
\end{equation}
where $\lambda_j^s= \varepsilon_{j}+s\bar \alpha_{j}k$, with $\bar\alpha_1 =\alpha_1$, $\bar\alpha_2 = t \alpha_2$, and the sign $t = \pm$ on $\alpha_2$ refers to the conventional and unconventional cases, respectively. Furthermore, $\beta_s = s\eta k$ and $\beta_s' =s \eta^* k$.
In the S basis, the eigenvalues and projection operators are easily found to be 
\begin{equation}
E_{sn}= \frac{1}{2}\left(\lambda_1^s+\lambda_2^s+n\sqrt{(\lambda_1^s-\lambda_2^s)^2+4\beta_s\beta_s'}\right), \
P^S_{sn}=\frac{1}{N_{sn}}
\begin{pmatrix}
  |\beta_s|^2 &\beta_s\delta^*_{sn}  \\
  \beta_s^*\delta_{sn}&|\delta_{sn}|^2
\end{pmatrix}, \  n=\pm 1,
\label{eigenvalues}
\end{equation}
where $\delta_{sn}=E_{sn}-\lambda_1^s$ and $N_{sn}^2=|\beta_s|^2+|\delta_{sn}|^2$. In the above, the index $n=\pm 1$ labels the eigenvalues for each block $s$. By transforming back the projection operators $P_{sn}^S$ in the original basis one finds 
\begin{equation}
P_{sn} (\theta ) =S^{-1}(\theta) P^S_{sn}S(\theta)=\frac{1}{2N_{sn}^2}
\begin{pmatrix}
    |\beta_s|^2 & -i s e^{-i\theta}|\beta_s|^2 & \beta_s\delta_{sn}^* &-i ts e^{-i\theta} \beta_s\delta_{sn}^* \\
 i s e^{i\theta}|\beta_s|^2     & |\beta_s|^2 &i s e^{i\theta} \beta_s\delta_{sn}^*  &t \beta_s\delta_{sn}^*  \\
 \beta_s^*\delta_{sn}     &-i s e^{-i\theta} \beta_s\delta_{sn}^* & |\delta_{sn}|^2 & -i  t s e^{-i\theta}|\beta_s|^2\\
i ts e^{i\theta} \beta_s\delta_{sn}^*       & t \beta_s^*\delta_{sn}  &i  t s e^{i\theta}|\beta_s|^2   &   |\delta_{sn}|^2 
\end{pmatrix}.
\label{projector_structure_first}
\end{equation}
Because the eigenvalues $E_{sn}$, as shown in Eq.~\eqref{eigenvalues}, only depend on the absolute value of ${\bf k}$, when the projector $P_{sn}(\theta )$ is inserted into Eq.~\eqref{selfenergy} for the self-energy, the latter acquires the simple structure 
\begin{equation}
    \Sigma^R (\omega, {\bf k})=-i \pi u_0^2 \sum_{s n} D_{s n} \langle P_{s n} (\theta )\rangle\equiv -i \pi u_0^2 \sum_{s n} D_{s n} P^0_{sn}(k_{sn}), 
    \label{self_energy_structure}
\end{equation}
where $\langle P_{s n} (\theta ) \rangle =P^0_{sn}(k_{sn})$ denotes the angle average and 
$D_{s n}$ is the density of states in the band with quantum numbers $s,n$, which reads
\begin{equation}
    D_{s n}= \frac{k }{2\pi }\frac{d k}{d E_{s n}(k)}\Big|_{k=k_{sn}(\mu)},
\end{equation}
$\mu$ being the Fermi energy, and $k_{sn}\equiv k_{sn}(\mu)$ is the Fermi momentum for each band.
We notice that the summation over $s,n$ actually runs only over the occupied subbands. When taking the angle average of the projector $P_{sn} (\theta )$ the self-energy reduces to only three independent parameters for the conventional and unconventional cases, greatly simplifying the numerical evaluation, and yields
\begin{align}
    \Sigma^R &\approx
    \begin{pmatrix}
        \Sigma_{11} & 0 & \Sigma_{12} & 0\\
        0 & \Sigma_{11} & 0 & t \Sigma_{12}\\
        \Sigma_{21} & 0 & \Sigma_{22} & 0\\
        0 & t\Sigma_{21} & 0 & \Sigma_{22}
    \end{pmatrix},
    \label{eq:Sigma}
\end{align}
where the sign $t = \pm$ refer to the conventional and unconventional cases.
In the first Born approximation, the Hermitian part of $\Sigma^R$ vanishes and $\Sigma^R$ becomes pure skew-symmetric, thus, it does not introduce renormalizations to $H$. Namely, we get
$\Re\Sigma_{11} = \Re\Sigma_{22} = 0$ in all cases. For the conventional case, $\Re\Sigma_{12} = \Re\Sigma_{21} = 0$ and $\Im\Sigma_{12}=\Im\Sigma_{21}$. For the unconventional case $\Im\Sigma_{12} = \Im\Sigma_{21} = 0$, and $\Re\Sigma_{12} = -\Re\Sigma_{21}$.
Although the above matrix structure has been obtained at the level of the Born approximation, it actually has a general validity granted by the symmetry properties of the model as will be shown in the following.

\begin{figure}[ht!]
    \centering
    \includegraphics[width=\columnwidth]{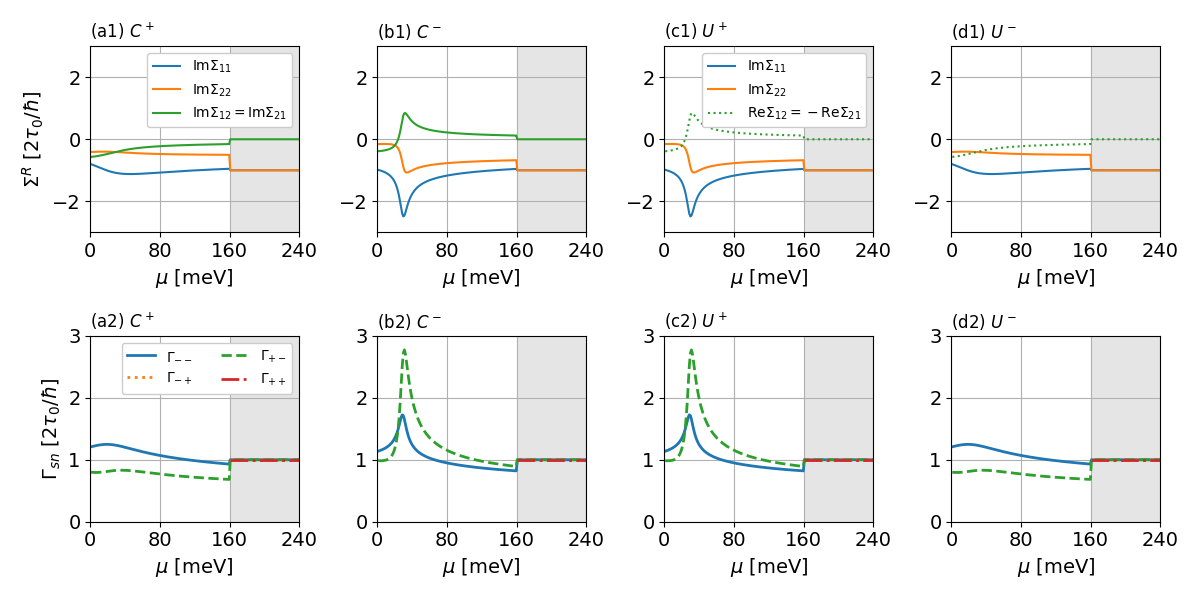}
    \caption{(a1-d1) Self-energy components [i.e., $\Sigma_{11}$, $\Sigma_{22}$, $\Sigma_{12}$, see Eq.~\eqref{eq:Sigma}] as function of the Fermi energy $\mu$. Each panel corresponds to the conventional ($C^\alpha$) or unconventional ($U^\alpha$) cases, as indicated, and $\alpha = \pm$ labels the relative sign of $\alpha_1/\alpha_2$.
    The gray area for $\mu > 160$~meV indicates the region where all subbands are occupied.
    (a2-d2) Imaginary part $\Gamma_{sn} = -\Im{E_{sn}}$ of the self-energy in the eigenstates basis considering only occupied subbands for each $\mu$.}
    \label{plots:self_eneergy}
\end{figure}

In Fig.~\ref{plots:self_eneergy} we show the behavior of the self-energy as a function of the 
Fermi energy. The self-energy is given in units of $\hbar/2\tau_0 = \pi u_0^2 D_0$, $D_0$ being the density of states of the two-dimensional quadratic dispersion.
Hence, $\pi u_0^2 D_0$ is the self-energy for the two-dimensional electron gas. In Fig.~\ref{plots:self_eneergy} one sees that, when all bands are occupied, the self-energy reduces to a unit matrix with $\Sigma_{11} = \Sigma_{22}$ and $\Sigma_{12}=0$. When only one band is occupied all three parameters are different from zero. In particular the self-energy in the empty subband is non zero due to the mixing of the inter-band SOC.

\subsection{The equation for the dressed vertex}

In this subsection we solve the vertex equation \eqref{vertex_equation} by using the {\it dressed} Green's function with the self-energy obtained in the previous section. We begin by introducing the {\it bare} current vertex 
\begin{equation}
    J^0_x ({\bf k})=\frac{1}{\hbar}\frac{\partial H({\bf k})}{\partial k_x}=\frac{\hbar k_x}{m}+J_{SOC,x},
    \label{bare_charge_current_x}
\end{equation}
where $J_{SOC,x}$ is a  ${\bf k}$-independent matrix arising from the linear-in-momentum terms describing the SOC in the Hamiltonian from Eq.~\eqref{generic_hamiltonian}. To solve Eq.~\eqref{vertex_equation} one proceeds by iteration i.e., perturbatively in the disorder correlations represented by dashed lines in the diagrams of Fig.~\ref{fig:diagrams}. In the first step of iteration, when the bare vertex from Eq.~\eqref{bare_charge_current_x} (the gray dot in Fig.~\ref{fig:diagrams} (c)) is inserted in the momentum integral over ${\bf k}'$ in the right hand side of Eq.~\eqref{vertex_equation}, one obtains the first vertex correction, which turns out to be independent on the external momentum ${\bf k}$. This suggests that a general solution must have the form
\begin{equation}
    J_x ({\bf k}) =  \frac{\hbar k_x}{m}+\Gamma_x,
     \label{ansatz_vertex}
\end{equation}
where $\Gamma_x$ is a momentum independent matrix.
By using the ansatz from Eq.~\eqref{ansatz_vertex} into the vertex equation \eqref{vertex_equation}, one readily obtains an {\it algebraic } equation for $\Gamma_x$
\begin{equation}
    \Gamma_x ={\bar\Gamma}_x
    +u_0^2  \int  \frac{d^2k'}{(2\pi)^2}G^R(\omega ,{\bf k}')
    \Gamma_x ({\bf k}')G^A (\omega, {\bf k}'),
    \label{algebraic_vertex_equation}
\end{equation}
where ${\bar \Gamma}_x$ is an {\it effective} bare vertex obtained by combining the {\it pure} SOC-induced vertex and the single-line impurity line dressed ordinary current vertex
\begin{equation}
    {\bar \Gamma}_x =J_{SOC,x}+u_0^2  \int  \frac{d^2k'}{(2\pi)^2}G^R(\omega ,{\bf k}')\frac{\hbar k_x}{m}G^A (\omega, {\bf k}').
    \label{bare_effective_vertex_x}
\end{equation}
The algebraic structure of Eq.~(\ref{algebraic_vertex_equation}) can be made explicit by introducing the basis $\zeta_a =\lambda_{a_1}\otimes \sigma_{a_2}$ (subband, spin) in the space of four by four matrices. The indices $(a_1, a_2)$ are defined as $a_{1}=\lfloor a/4\rfloor$, $\lfloor \cdots\rfloor$ being the integer part, and $a_2=a$ ${\rm Mod} \ 4$.
Table \ref{tab:index_zeta_matrices} provides all the indices $a$ for each pair $(\lambda, \sigma)$.
By introducing the decompositions $\Gamma_x =\sum_a \Gamma_{x,a}\zeta_a$ and similarly for ${\bar \Gamma}_x$, one finds
\begin{equation}
    \sum_b
    \left( \delta _{ab}-L_{ab}\right)\Gamma_{x,b}={\bar \Gamma}_{x,a},
    \label{linear_vertex_system}
\end{equation}
with the $L$-matrix given by
\begin{equation}
    L_{ab}=u_0^2  \int  \frac{d^2k'}{(2\pi)^2}
    \dfrac{1}{4}
    {\rm Tr}\left[\zeta_a \  G^R(\omega ,{\bf k}')\zeta_b \  G^A (\omega, {\bf k}')\right].
    \label{L-matrix}
\end{equation}

\begin{table}[ht]
    \centering
    \begin{tabular}{|c||c|c|c|c|}\hline
       $(\lambda , \sigma)$  & 0&1&2&3 \\ \hline \hline
        0 & 0&1&2&3 \\ \hline
         1 & 4&5&6&7 \\ \hline
          2 & 8&9&10&11 \\ \hline 
           3 & 12&13&14&15 \\ \hline 
    \end{tabular}
    \vskip 0.5cm
    \caption{For each pair of $(\lambda, \sigma)$ matrices, the table gives the value of the index $a$.}
    \label{tab:index_zeta_matrices}
\end{table}

Given the four-dimensional structure of the Hamiltonian  from Eq.~\eqref{generic_hamiltonian}, the basis $\zeta_a$ is formed by   sixteen matrices, making a purely analytical approach to Eq.~\eqref{linear_vertex_system} almost impossible. However, even though a numerical approach is possible, it is useful to investigate how the symmetries of the model allow to reduce the algebraic system of the vertex equations in a set of independent  systems of equations of lesser size.
Because each of the matrix $\zeta_a$ can be connected to an observable, such a reduction allows also to elucidate which physical observables are connected to one another.

\subsection{The symmetry analysis}

To illustrate the usefulness of the symmetry analysis, we begin by considering the evaluation of the effective bare vertex ${\bar \Gamma}_x$ defined in Eq.~\eqref{bare_effective_vertex_x}. The $\zeta$-matrices decomposition of the matrix part $J_{SOC, x}$ is easily obtained and reads
\begin{equation}
    J_{SOC, x}=\frac{\alpha_1+\alpha_2}{2\hbar}\zeta_2 - \frac{\eta_U}{\hbar}\zeta_5 +\frac{\eta_C}{\hbar}\zeta_6 +\frac{\alpha_1-\alpha_2}{2\hbar}\zeta_{14}.
    \label{zeta_decomposition_JSOC}
\end{equation}
Notice that by restricting to the cases $\alpha_1 =\pm \alpha_2$, the matrix part of  $J_{SOC, x}$ has only two components associated to intra- and interband SOC.

In order to obtain the full $\bar{\Gamma}_x$ we need to evaluate the momentum integral of the second term in the right hand side of Eq.~\eqref{bare_effective_vertex_x}. After expressing the Green's functions in terms of their spectral decomposition, one gets for that integral
\begin{equation}
    u_0^2\sum_{s  n } \int \frac{d^2 k}{(2\pi)^2} \
    \frac{k_x P_{sn}({\bf k}) }{(\hbar \omega -E_{sn}({\bf k})+i0^+)(\hbar \omega -E_{sn}({\bf k})-i0^+)}, 
    \label{integral_single_impurity_vertex}
\end{equation}
where the orthogonality of the projectors has been used
$P_{sn}({\bf k}) P_{s'n'} ({\bf k})=\delta_{ss'}\delta_{nn'} P_{sn}({\bf k})$.
According to the explicit expression of the projector in Eq.~\eqref{projector_structure_first}, we may represent the angle dependence of $P_{sn}({\bf k})$ as
\begin{equation}
    P_{sn}({\bf k}) = P^0_{sn}({ k})+e^{i\theta}P^+_{sn}({ k})+e^{-i\theta}P^-_{sn}({ k})
    \label{angle_dependence_projector}.
\end{equation}
The angle integration in Eq.~\eqref{integral_single_impurity_vertex}, due to the $\cos \theta$ factor of $k_x$, yields $P^0_{sn}(k)$ and
$P^1_{sn}(k)=(1/2) (P^+_{sn}({ k})+P^-_{sn}({ k}))$, whose $\zeta$-matrices decomposition is readily obtained by inspection looking at Eq.~\eqref{projector_structure_first}, yielding
\begin{align}
     P^0_{sn}(k) &= \frac{1}{2N_s^2}
     \left(\frac{|\beta_s|^2+|\delta_{sn}|^2}{2} \zeta_0 +\frac{|\beta_s|^2-|\delta_{sn}|^2}{2}\zeta_{12} +\frac{\beta_n \delta^*_{sn}+ \beta^*_n \delta_{sn}}{2} \zeta_{4} +
     i\frac{\beta_n \delta^*_{sn}- \beta^*_n \delta_{sn}}{2}
     \zeta_{8}\right)
     \label{angle_P0_integrated_projector_C}
\\
    P^1_{sn}(k) &= \frac{1}{4N_s^2}
    s\left( \frac{|\beta_s|^2+|\delta_{sn}|^2}{2} \zeta_2+ \frac{|\beta_s|^2-|\delta_{sn}|^2}{2}
    \zeta_{14} 
    +\beta_n \delta^*_{sn}(\zeta_{6}+i\zeta_{10})-
    \beta^*_n \delta_{sn} (\zeta_{6}-i\zeta_{10})\right)
    \label{angle_integrated_projector_C}
\end{align}
in the conventional case and 
\begin{align}
    P^0_{sn}(k)  &=\frac{1}{2N_s^2}
    \left(\frac{|\beta_s|^2+|\delta_{sn}|^2}{2} \zeta_0 +\frac{|\beta_s|^2-|\delta_{sn}|^2}{2}\zeta_{12} +\frac{\beta_n \delta^*_{sn}+ \beta^*_n \delta_{sn}}{2} \zeta_{7} +
    i\frac{\beta_n \delta^*_{sn}- \beta^*_n \delta_{sn}}{2}
    \zeta_{11}\right)
    \label{angle_P0_integrated_projector_U}
\\
    P^1_{sn}(k)  &= \frac{1}{4N_s^2}
    s\left( \frac{|\beta_s|^2+|\delta_{sn}|^2}{2} \zeta_{14}+ \frac{|\beta_s|^2-|\delta_{sn}|^2}{2}
    \zeta_{2} 
    +\beta_n \delta^*_{sn}(i\zeta_{5}-\zeta_{9})-
    \beta^*_n \delta_{sn} (-i\zeta_{5}+\zeta_{9})\right)
    \label{angle_integrated_projector_U}
\end{align}
for the unconventional case. By comparing Eq.~\eqref{zeta_decomposition_JSOC} with Eqs.~(\ref{angle_integrated_projector_C}, \ref{angle_integrated_projector_U}) one sees that only the sets $(\zeta_2, \zeta_{14}, \zeta_6, \zeta_{10})$ and $(\zeta_2, \zeta_{14}, \zeta_5, \zeta_9)$ are involved in the conventional and unconventional cases, respectively.
This suggests that in the evaluation of the full vertex $\Gamma_x$ one has to deal with an algebraic system of dimension four, which is a great simplification with respect to the dimension sixteen expected on general grounds. 
Finally, by looking at Eqs.~(\ref{angle_P0_integrated_projector_C}, \ref{angle_P0_integrated_projector_U}) one sees that it involves only the sets $(\zeta_0, \zeta_{12}, \zeta_{4}, \zeta_{8})$ and $(\zeta_0, \zeta_{12}, \zeta_{7}, \zeta_{11})$ for the conventional and unconventional cases, respectively. These are precisely the sets that determine the most general structure of the self-energy obtained in Eq.~\eqref{self_energy_structure}.

We now show that the above reduction of the observables in independent sets is dictated by the symmetry properties of the model.

\subsubsection{The conventional SOC case}
In the conventional case, the symmetry group of the model is $C_{2V}$ with generators given by 
twofold rotations about the z axis $C_2(z)= -i\lambda_0\otimes R_2 (z)$, mirror reflection through the y axis  $M_y= -\lambda_0\otimes R_2(y)$ and time reversal $T=-i\lambda_0 \otimes \sigma_y K$, where $R_n ({\hat u})=\exp (i (\pi/n) {\hat u}\cdot {\boldsymbol \sigma})$ is the n-fold spin rotation around the axis given by the unit vector ${\hat u}$ and $K$ is complex conjugation. Under any of these symmetry operations, each $\zeta_a$ matrix transforms as $\zeta_a\rightarrow \pm \zeta_a$. It is then possible to derive the parity eigenvalues for each $\zeta_a$ matrix and the result is shown in Table \ref{paritytables}. We see that the observables $\zeta_a$ divide in four groups referring to the charge and the different spin polarization degrees of freedom. One may see that the third group, corresponding to the $y$-axis spin polarization includes exactly the set of $\zeta_a$ matrices identified in the analysis of the single-impurity line diagram analyzed in Eq.~\eqref{angle_integrated_projector_C}.


\subsubsection{The unconventional SOC case}
In the unconventional case, the symmetry group of the model is $C_{3V}$ with generators given by 
threefold rotations about the z axis $C_3(z)= -i\lambda_0\otimes R_3 (z)$, mirror reflection through the y axis  $M_y= \lambda_z\otimes i \sigma_y$ and time reversal $T=\lambda_0 \otimes i \sigma_y K$. In this case, under  the $M_y$ and $T$ symmetry operations above, each $\zeta_a$ matrix transforms as $\zeta_a\rightarrow \pm \zeta_a$ and we obtain the  parity eigenvalues table which  is shown in Table \ref{paritytables}. Under the $C_{3}(z)$ symmetry operations, instead, the $\zeta_a$ matrices with spin polarization along the x and y axis transform into one another and do not have a defined parity. Nonetheless, as it is clear from the Table \ref{paritytables}, the $\zeta_a$ matrices split again in four distinct groups. The last two groups are related to the x and y axes spin polarization, although in a manner completely different from the conventional case. Indeed, the intraband observables (with the band index $\lambda =0,3$) correspond to the given spin polarization components, whereas the interband ones (with the band index $\lambda =1,2$) correspond to the other in-plane spin component, responsible for spin admixture. Similarly, the first two groups in Table \ref{paritytables} describe the charge and z-axis spin polarization, again showing a different symmetry in intra- and interband terms.
Finally, as in the case of the conventional case, the set $(\zeta_{2},\zeta_{5}, \zeta_{14}, \zeta_{9})$, which is the third group in Table  \ref{paritytables}, corresponds to the set which appears in  the single-impurity diagram for the effective bare vertex found in Eq.~\eqref{angle_integrated_projector_U}.

\begin{table}[ht]
\centering
\begin{tabular}{cl|cccccc|c|cccccc|}
\cline{3-8} \cline{10-15}
                                                                  &  & \multicolumn{6}{c|}{Conventional: $C_{2V}$}                                                                                                                                                                                                                                                          &  & \multicolumn{6}{c|}{Unconventional: $C_{3V}$}                                                                                                                                                                                                                                                                                                             \\ \cline{1-1} \cline{3-8} \cline{10-15} 
\multicolumn{1}{|c|}{Set}                                       &  & \multicolumn{1}{c|}{$a$}                        & \multicolumn{1}{c|}{$(\lambda,\sigma)$}            & \multicolumn{1}{c|}{$C_2(z)$}                  & \multicolumn{1}{c|}{$M_y$}                     & \multicolumn{1}{c|}{$T$}                       & Class                            &  & \multicolumn{1}{c|}{$a$}                        & \multicolumn{1}{c|}{$(\lambda,\sigma)$}            & \multicolumn{1}{c|}{$C_3(z)$}                                                                      & \multicolumn{1}{c|}{$M_y$}                     & \multicolumn{1}{c|}{$T$}                       & Class                             \\ \cline{1-1} \cline{3-8} \cline{10-15} 
\multicolumn{1}{|c|}{\cellcolor[HTML]{EFEFEF}}                    &  & \multicolumn{1}{c|}{\cellcolor[HTML]{EFEFEF}0}  & \multicolumn{1}{c|}{\cellcolor[HTML]{EFEFEF}(0,0)} & \multicolumn{1}{c|}{\cellcolor[HTML]{EFEFEF}+} & \multicolumn{1}{c|}{\cellcolor[HTML]{EFEFEF}+} & \multicolumn{1}{c|}{\cellcolor[HTML]{EFEFEF}+} & \cellcolor[HTML]{EFEFEF}$A_1(s)$ &  & \multicolumn{1}{c|}{\cellcolor[HTML]{EFEFEF}0}  & \multicolumn{1}{c|}{\cellcolor[HTML]{EFEFEF}(0,0)} & \multicolumn{1}{c|}{\cellcolor[HTML]{EFEFEF}+}                                                     & \multicolumn{1}{c|}{\cellcolor[HTML]{EFEFEF}+} & \multicolumn{1}{c|}{\cellcolor[HTML]{EFEFEF}+} & \cellcolor[HTML]{EFEFEF}$A_1 (s)$ \\ \cline{3-8} \cline{10-15} 
\multicolumn{1}{|c|}{\cellcolor[HTML]{EFEFEF}}                    &  & \multicolumn{1}{c|}{\cellcolor[HTML]{EFEFEF}4}  & \multicolumn{1}{c|}{\cellcolor[HTML]{EFEFEF}(1,0)} & \multicolumn{1}{c|}{\cellcolor[HTML]{EFEFEF}+} & \multicolumn{1}{c|}{\cellcolor[HTML]{EFEFEF}+} & \multicolumn{1}{c|}{\cellcolor[HTML]{EFEFEF}+} & \cellcolor[HTML]{EFEFEF}$A_1(s)$ &  & \multicolumn{1}{c|}{\cellcolor[HTML]{EFEFEF}11} & \multicolumn{1}{c|}{\cellcolor[HTML]{EFEFEF}(2,3)} & \multicolumn{1}{c|}{\cellcolor[HTML]{EFEFEF}+}                                                     & \multicolumn{1}{c|}{\cellcolor[HTML]{EFEFEF}+} & \multicolumn{1}{c|}{\cellcolor[HTML]{EFEFEF}+} & \cellcolor[HTML]{EFEFEF}$A_1 (s)$ \\ \cline{3-8} \cline{10-15} 
\multicolumn{1}{|c|}{\cellcolor[HTML]{EFEFEF}}                    &  & \multicolumn{1}{c|}{\cellcolor[HTML]{EFEFEF}12} & \multicolumn{1}{c|}{\cellcolor[HTML]{EFEFEF}(3,0)} & \multicolumn{1}{c|}{\cellcolor[HTML]{EFEFEF}+} & \multicolumn{1}{c|}{\cellcolor[HTML]{EFEFEF}+} & \multicolumn{1}{c|}{\cellcolor[HTML]{EFEFEF}+} & \cellcolor[HTML]{EFEFEF}$A_1(s)$ &  & \multicolumn{1}{c|}{\cellcolor[HTML]{EFEFEF}12} & \multicolumn{1}{c|}{\cellcolor[HTML]{EFEFEF}(3,0)} & \multicolumn{1}{c|}{\cellcolor[HTML]{EFEFEF}+}                                                     & \multicolumn{1}{c|}{\cellcolor[HTML]{EFEFEF}+} & \multicolumn{1}{c|}{\cellcolor[HTML]{EFEFEF}+} & \cellcolor[HTML]{EFEFEF}$A_1 (s)$ \\ \cline{3-8} \cline{10-15} 
\multicolumn{1}{|c|}{\multirow{-4}{*}{\cellcolor[HTML]{EFEFEF}1}} &  & \multicolumn{1}{c|}{\cellcolor[HTML]{EFEFEF}8}  & \multicolumn{1}{c|}{\cellcolor[HTML]{EFEFEF}(2,0)} & \multicolumn{1}{c|}{\cellcolor[HTML]{EFEFEF}+} & \multicolumn{1}{c|}{\cellcolor[HTML]{EFEFEF}+} & \multicolumn{1}{c|}{\cellcolor[HTML]{EFEFEF}-} & \cellcolor[HTML]{EFEFEF}$A_1(s)$ &  & \multicolumn{1}{c|}{\cellcolor[HTML]{EFEFEF}7}  & \multicolumn{1}{c|}{\cellcolor[HTML]{EFEFEF}(1,3)} & \multicolumn{1}{c|}{\cellcolor[HTML]{EFEFEF}+}                                                     & \multicolumn{1}{c|}{\cellcolor[HTML]{EFEFEF}+} & \multicolumn{1}{c|}{\cellcolor[HTML]{EFEFEF}-} & \cellcolor[HTML]{EFEFEF}$A_1 (s)$ \\ \cline{1-1} \cline{3-8} \cline{10-15} 
\multicolumn{1}{|c|}{}                                            &  & \multicolumn{1}{c|}{3}                          & \multicolumn{1}{c|}{(0,3)}                         & \multicolumn{1}{c|}{+}                         & \multicolumn{1}{c|}{-}                         & \multicolumn{1}{c|}{-}                         & $A_2(R_z)$                       &  & \multicolumn{1}{c|}{3}                          & \multicolumn{1}{c|}{(0,3)}                         & \multicolumn{1}{c|}{+}                                                                             & \multicolumn{1}{c|}{-}                         & \multicolumn{1}{c|}{-}                         & $A_2 (R_z)$                       \\ \cline{3-8} \cline{10-15} 
\multicolumn{1}{|c|}{}                                            &  & \multicolumn{1}{c|}{7}                          & \multicolumn{1}{c|}{(1,3)}                         & \multicolumn{1}{c|}{+}                         & \multicolumn{1}{c|}{-}                         & \multicolumn{1}{c|}{-}                         & $A_2(R_z)$                       &  & \multicolumn{1}{c|}{8}                          & \multicolumn{1}{c|}{(2,0)}                         & \multicolumn{1}{c|}{+}                                                                             & \multicolumn{1}{c|}{-}                         & \multicolumn{1}{c|}{-}                         & $A_2 (R_z)$                       \\ \cline{3-8} \cline{10-15} 
\multicolumn{1}{|c|}{}                                            &  & \multicolumn{1}{c|}{15}                         & \multicolumn{1}{c|}{(3,3)}                         & \multicolumn{1}{c|}{+}                         & \multicolumn{1}{c|}{-}                         & \multicolumn{1}{c|}{-}                         & $A_2(R_z)$                       &  & \multicolumn{1}{c|}{15}                         & \multicolumn{1}{c|}{(3,3)}                         & \multicolumn{1}{c|}{+}                                                                             & \multicolumn{1}{c|}{-}                         & \multicolumn{1}{c|}{-}                         & $A_2 (R_z)$                       \\ \cline{3-8} \cline{10-15} 
\multicolumn{1}{|c|}{\multirow{-4}{*}{2}}                         &  & \multicolumn{1}{c|}{11}                         & \multicolumn{1}{c|}{(2,3)}                         & \multicolumn{1}{c|}{+}                         & \multicolumn{1}{c|}{-}                         & \multicolumn{1}{c|}{+}                         & $A_2(R_z)$                       &  & \multicolumn{1}{c|}{4}                          & \multicolumn{1}{c|}{(1,0)}                         & \multicolumn{1}{c|}{+}                                                                             & \multicolumn{1}{c|}{-}                         & \multicolumn{1}{c|}{+}                         & $A_2 (R_z)$                       \\ \cline{1-1} \cline{3-8} \cline{10-15} 
\multicolumn{1}{|c|}{\cellcolor[HTML]{EFEFEF}}                    &  & \multicolumn{1}{c|}{\cellcolor[HTML]{EFEFEF}2}  & \multicolumn{1}{c|}{\cellcolor[HTML]{EFEFEF}(0,2)} & \multicolumn{1}{c|}{\cellcolor[HTML]{EFEFEF}-} & \multicolumn{1}{c|}{\cellcolor[HTML]{EFEFEF}+} & \multicolumn{1}{c|}{\cellcolor[HTML]{EFEFEF}-} & \cellcolor[HTML]{EFEFEF}$B_1(x)$ &  & \multicolumn{1}{c|}{\cellcolor[HTML]{EFEFEF}2}  & \multicolumn{1}{c|}{\cellcolor[HTML]{EFEFEF}(0,2)} & \multicolumn{1}{c|}{\cellcolor[HTML]{EFEFEF}$-\frac{\sqrt{3}}{2}\zeta_1-\frac{1}{2}\zeta_2$}       & \multicolumn{1}{c|}{\cellcolor[HTML]{EFEFEF}+} & \multicolumn{1}{c|}{\cellcolor[HTML]{EFEFEF}-} & \cellcolor[HTML]{EFEFEF}$E(x,y)$  \\ \cline{3-8} \cline{10-15} 
\multicolumn{1}{|c|}{\cellcolor[HTML]{EFEFEF}}                    &  & \multicolumn{1}{c|}{\cellcolor[HTML]{EFEFEF}6}  & \multicolumn{1}{c|}{\cellcolor[HTML]{EFEFEF}(1,2)} & \multicolumn{1}{c|}{\cellcolor[HTML]{EFEFEF}-} & \multicolumn{1}{c|}{\cellcolor[HTML]{EFEFEF}+} & \multicolumn{1}{c|}{\cellcolor[HTML]{EFEFEF}-} & \cellcolor[HTML]{EFEFEF}$B_1(x)$ &  & \multicolumn{1}{c|}{\cellcolor[HTML]{EFEFEF}5}  & \multicolumn{1}{c|}{\cellcolor[HTML]{EFEFEF}(1,1)} & \multicolumn{1}{c|}{\cellcolor[HTML]{EFEFEF}$+\frac{\sqrt{3}}{2}\zeta_6-\frac{1}{2}\zeta_5$}       & \multicolumn{1}{c|}{\cellcolor[HTML]{EFEFEF}+} & \multicolumn{1}{c|}{\cellcolor[HTML]{EFEFEF}-} & \cellcolor[HTML]{EFEFEF}$E(x,y)$  \\ \cline{3-8} \cline{10-15} 
\multicolumn{1}{|c|}{\cellcolor[HTML]{EFEFEF}}                    &  & \multicolumn{1}{c|}{\cellcolor[HTML]{EFEFEF}14} & \multicolumn{1}{c|}{\cellcolor[HTML]{EFEFEF}(3,2)} & \multicolumn{1}{c|}{\cellcolor[HTML]{EFEFEF}-} & \multicolumn{1}{c|}{\cellcolor[HTML]{EFEFEF}+} & \multicolumn{1}{c|}{\cellcolor[HTML]{EFEFEF}-} & \cellcolor[HTML]{EFEFEF}$B_1(x)$ &  & \multicolumn{1}{c|}{\cellcolor[HTML]{EFEFEF}14} & \multicolumn{1}{c|}{\cellcolor[HTML]{EFEFEF}(3,2)} & \multicolumn{1}{c|}{\cellcolor[HTML]{EFEFEF}$-\frac{\sqrt{3}}{2}\zeta_{13}-\frac{1}{2}\zeta_{14}$} & \multicolumn{1}{c|}{\cellcolor[HTML]{EFEFEF}+} & \multicolumn{1}{c|}{\cellcolor[HTML]{EFEFEF}-} & \cellcolor[HTML]{EFEFEF}$E(x,y)$  \\ \cline{3-8} \cline{10-15} 
\multicolumn{1}{|c|}{\multirow{-4}{*}{\cellcolor[HTML]{EFEFEF}3}} &  & \multicolumn{1}{c|}{\cellcolor[HTML]{EFEFEF}10} & \multicolumn{1}{c|}{\cellcolor[HTML]{EFEFEF}(2,2)} & \multicolumn{1}{c|}{\cellcolor[HTML]{EFEFEF}-} & \multicolumn{1}{c|}{\cellcolor[HTML]{EFEFEF}+} & \multicolumn{1}{c|}{\cellcolor[HTML]{EFEFEF}+} & \cellcolor[HTML]{EFEFEF}$B_1(x)$ &  & \multicolumn{1}{c|}{\cellcolor[HTML]{EFEFEF}9}  & \multicolumn{1}{c|}{\cellcolor[HTML]{EFEFEF}(2,1)} & \multicolumn{1}{c|}{\cellcolor[HTML]{EFEFEF}$+\frac{\sqrt{3}}{2}\zeta_{10}-\frac{1}{2}\zeta_9$}    & \multicolumn{1}{c|}{\cellcolor[HTML]{EFEFEF}+} & \multicolumn{1}{c|}{\cellcolor[HTML]{EFEFEF}+} & \cellcolor[HTML]{EFEFEF}$E(x,y)$  \\ \cline{1-1} \cline{3-8} \cline{10-15} 
\multicolumn{1}{|c|}{}                                            &  & \multicolumn{1}{c|}{1}                          & \multicolumn{1}{c|}{(0,1)}                         & \multicolumn{1}{c|}{-}                         & \multicolumn{1}{c|}{-}                         & \multicolumn{1}{c|}{-}                         & $B_2(y)$                         &  & \multicolumn{1}{c|}{1}                          & \multicolumn{1}{c|}{(0,1)}                         & \multicolumn{1}{c|}{$+\frac{\sqrt{3}}{2}\zeta_2-\frac{1}{2}\zeta_1$}                               & \multicolumn{1}{c|}{-}                         & \multicolumn{1}{c|}{-}                         & \cellcolor[HTML]{EFEFEF}$E(x,y)$  \\ \cline{3-8} \cline{10-15} 
\multicolumn{1}{|c|}{}                                            &  & \multicolumn{1}{c|}{5}                          & \multicolumn{1}{c|}{(1,1)}                         & \multicolumn{1}{c|}{-}                         & \multicolumn{1}{c|}{-}                         & \multicolumn{1}{c|}{-}                         & $B_2(y)$                         &  & \multicolumn{1}{c|}{6}                          & \multicolumn{1}{c|}{(1,2)}                         & \multicolumn{1}{c|}{$-\frac{\sqrt{3}}{2}\zeta_5-\frac{1}{2}\zeta_6$}                               & \multicolumn{1}{c|}{-}                         & \multicolumn{1}{c|}{-}                         & \cellcolor[HTML]{EFEFEF}$E(x,y)$  \\ \cline{3-8} \cline{10-15} 
\multicolumn{1}{|c|}{}                                            &  & \multicolumn{1}{c|}{13}                         & \multicolumn{1}{c|}{(3,1)}                         & \multicolumn{1}{c|}{-}                         & \multicolumn{1}{c|}{-}                         & \multicolumn{1}{c|}{-}                         & $B_2(y)$                         &  & \multicolumn{1}{c|}{13}                         & \multicolumn{1}{c|}{(3,1)}                         & \multicolumn{1}{c|}{$+\frac{\sqrt{3}}{2}\zeta_{14}-\frac{1}{2}\zeta_{13}$}                         & \multicolumn{1}{c|}{-}                         & \multicolumn{1}{c|}{-}                         & \cellcolor[HTML]{EFEFEF}$E(x,y)$  \\ \cline{3-8} \cline{10-15} 
\multicolumn{1}{|c|}{\multirow{-4}{*}{4}}                         &  & \multicolumn{1}{c|}{9}                          & \multicolumn{1}{c|}{(2,1)}                         & \multicolumn{1}{c|}{-}                         & \multicolumn{1}{c|}{-}                         & \multicolumn{1}{c|}{+}                         & $B_2(y)$                         &  & \multicolumn{1}{c|}{10}                         & \multicolumn{1}{c|}{(2,2)}                         & \multicolumn{1}{c|}{$-\frac{\sqrt{3}}{2}\zeta_9-\frac{1}{2}\zeta_{10}$}                            & \multicolumn{1}{c|}{-}                         & \multicolumn{1}{c|}{+}                         & \cellcolor[HTML]{EFEFEF}$E(x,y)$  \\ \cline{1-1} \cline{3-8} \cline{10-15} 
\end{tabular}
\vspace{0.3cm}
\caption{
    Parity of the $\zeta_a$ matrices under the symmetry transformations for the conventional and unconventional Rashba models. For the conventional model ($C_{2V}$ group), the matrices divide into four sets corresponding to charge $\sigma =0$ and spin polarization along the i-th axis $\sigma =1$, $i=1,2,3$. The case of the $y$ polarization, $i=2$, selects the matrices $\zeta_2$, $\zeta_{6}$, $\zeta_{14}$ and $\zeta_{10}$.
    For the unconventional model ($C_{3V}$ group), the first and second groups mix charge ($\sigma=0$) and spin-z ($\sigma=3$) components, while the third and fourth groups mix in-plane spin components. The $C_3(z)$ symmetry partners of the irrep $E$ are split into the last two groups.
}
\label{paritytables}
\end{table}

\subsubsection{Analysis of the 
\texorpdfstring{$L_{ab}$}{Lab}
-matrix}

The solution of the vertex equation \eqref{algebraic_vertex_equation} is determined by the $L_{ab}$ matrix, whose entries are given by the momentum integral in Eq.~\eqref{L-matrix}. We notice that the $L_{ab}$ matrix is dimensionless and must be of order one, no matter what is the strength of disorder. Consider the case, for instance, when disorder is vanishingly small (i. $u_0^2\rightarrow 0$). One could naively assume that the $L_{ab}$ should vanish. This is not so because for vanishing disorder, the poles of the retarded and advanced Green's  functions tend to merge towards the real axis and the momentum integral diverges as $u_0^{-2}$, thus compensating exactly the factor $u_0^2$ in front of the integral. In this case the Green's functions become those for the clean system and one can analyze the symmetry properties by relying on the parity of the $\zeta_a$ matrices. 

Under a unitary symmetry transformation the Green's functions remain invariant, whereas the $\zeta_a$ matrices transforms accordingly to Table \ref{paritytables}. Thus, for the symmetry operations $S$ with well defined parities, the $L$ matrix transform as $L_{ab} \rightarrow p_a p_b L_{ab}$, where $p_a$ and $p_b$ are the parity eigenvalues determined in the Table \ref{paritytables}. 
It is clear that the entry $L_{ab}$ may only differ from zero if and only if $p_a p_b >0$, i.e. if the two $\zeta$ matrices transform with the same sign of parity. This implies, according to the analysis carried in the previous subsections, that the $L$ matrix decomposes into independent blocks. 
Additionally, under time-reversal symmetry $G^R \leftrightarrow G^A$, thus $L_{ab} \rightarrow p_a p_b L_{ba}$. Finally, the $C_3(z)$ symmetry introduces further constraints to the third and forth blocks from Table \ref{paritytables}, but this analysis is not necessary to identify the four split blocks.

\begin{figure}[ht!]
    \centering
    \includegraphics[width=\columnwidth]{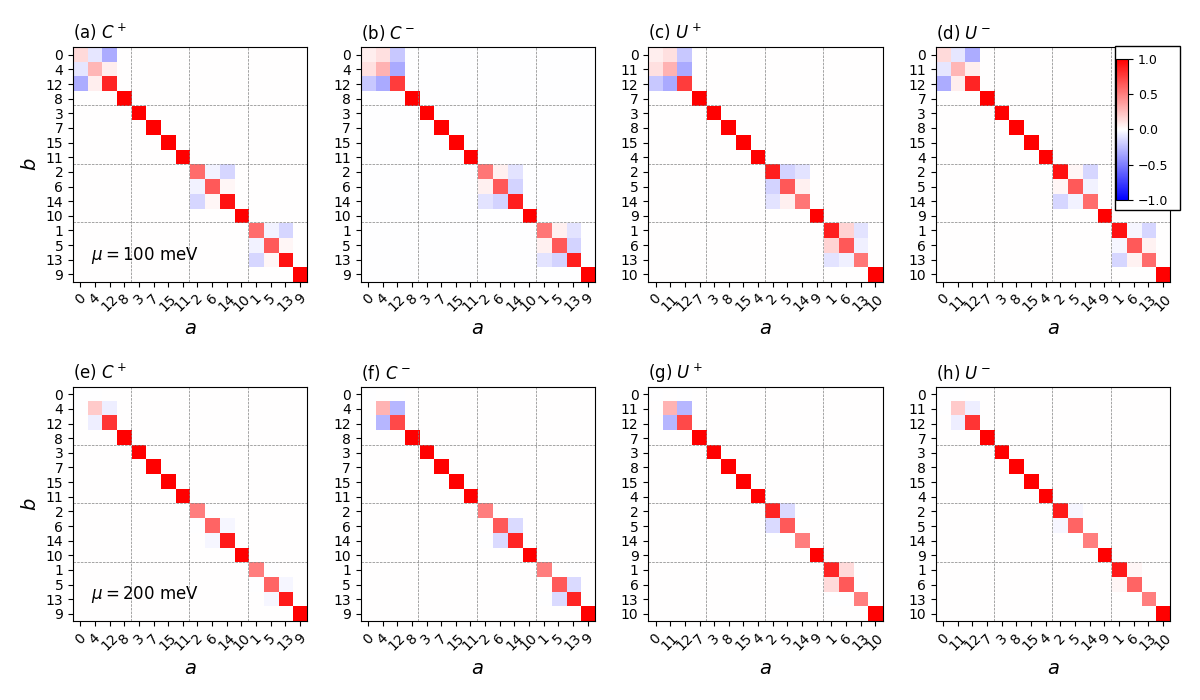}
    \caption{Entries of the matrix $\delta_{ab}-L_{a,b}$ with a color scale. (a-d) Top panels are calculated for $\mu=100$~meV, such that the chemical potential lies in the first subband and hence only one band is occupied. (e-h) For the bottom panels we use $\mu=200$~meV, such that both subbands are occupied. In the bottom row the white square at the $(0,0)$ position signal the vanishing eigenvalue associate to the total charge mode.}
    \label{L_blocks}
\end{figure}

The $L$ matrix or better the combination ${\bf 1} -L$, which appears in the vertex equation \eqref{linear_vertex_system} can be numerically evaluated. In Fig.~\ref{L_blocks} we plot the entries of the $L$ matrix using a color code to illustrate the block structure. We consider two typical and distinct cases. In Figs.~\ref{L_blocks}(e-h) the chemical potential lies in the second subband and both subbands are occupied, whereas in Figs.~\ref{L_blocks}(a-d) only the lowest subband is occupied. The numbers on the axes refer to the $a$ and $b$ indices of the $\zeta$ matrices. Clearly the blocks confirm the symmetry analysis carried out before.

The symmetry analysis of the $L$-matrix can be connected to  the general structure of the self-energy we obtained in Eq.~\eqref{eq:Sigma}. By using the identity\cite{schwab:2002} $G^R-G^A=G^R (\Sigma^R -\Sigma^A)G^A$ in Eq.~\eqref{selfenergy} we get
\begin{equation}
 \Sigma^R-\Sigma^A= u_0^2 \int \frac{d^2k'}{(2\pi)^2}    G^R(\omega, {\bf k'}) (\Sigma^R -\Sigma^A )G^A(\omega, {\bf k'}).
\end{equation}
Since the self-energy does not depend on the momentum, the above equation can be transformed into an algebraic one as
\begin{equation}
 ( \Sigma^R-\Sigma^A )_a =L_{ab} ( \Sigma^R-\Sigma^A )_b.
 \label{vector_scattering_rates}
\end{equation}
This shows that the skew-Hermitian part of the self-energy, which forms a vector of scattering rates with components $a=0, \dots , 15$, has a  vanishing eigenvalue for the matrix ${\bf 1}-L$ discussed earlier, i.e. it belongs to the nullspace of ${\bf 1}-L$. 
When the self-energy is proportional to the identity matrix $\zeta_0$ the null space reduces to the total charge sector and the vanishing eigenvalue of ${\bf 1}-L$
is the manifestation of charge conservation. This is what happens in the absence of SOC.
In the presence of SOC, by considering Fig.~\ref{L_blocks}, we may clearly distinguish the two different regimes corresponding to one (upper row, energy $\mu =100$ meV) or two (lower row, energy $\mu =200$ meV)
bands occupied. When only one band is occupied there is a coupling among the set $\zeta_0$, $\zeta_4$ and $\zeta_{12}$ for the conventional model, or set $\zeta_0$, $\zeta_{11}$ and $\zeta_{12}$ for the unconventional models. This implies that the null space of the ${\bf 1}-L$ has components in the $\zeta$ matrices of the above sets, i.e. the total charge fluctuating mode is coupled to the charge transfer modes between the subbands. In the case of two bands occupied  the total charge sector decouples from the other   charge transfer modes and has a zero eigenvalue  as shown in the color code of  Fig.~\ref{L_blocks}. Indeed, this is what is expected based on the self-energy evaluation when the latter reduces to $\zeta_0$.
The two charge transfer modes in this regime remain coupled among themselves.  This is evidenced in the small coloured square in the top left square in each panel of the bottom row in Fig.~\ref{L_blocks}.

 Another noticeable feature in Fig.~\ref{L_blocks} are the  diagonal blocks, which describe the in-plane spin modes. For instance, in the case of the conventional model,  the third diagonal block corresponding to the sets of matrices $\zeta_2, \zeta_{6}, \zeta_{14}, \zeta_{10}$ includes the spin mode with y-axis polarization.  One sees that in the case of only one band occupied (top row, panels (a) and (b)) the spin mode with y-axis polarization is coupled to the spin transfer mode between the bands with the same y-axis polarization.
On the other hand, when both bands are occupied, the spin mode with y-axis polarization decouples and behaves in a way similar to the single-band case. 

In the case of the unconventional model we focus again on third diagonal block with the matrices  $\zeta_2, \zeta_{5}, \zeta_{14}, \zeta_{9}$. In the regime when only one band is occupied, the spin mode with $y$-axis polarization is coupled to two spin transfer modes between the bands with both $y$-axis ($\zeta_{14}$)and the  x-axis polarization ($\zeta_5$), in sharp contrast with the conventional case. Even when both bands are occupied  
the spin mode with $y$-axis polarization remains coupled with the mode with   $x$-axis polarization ($\zeta_5$).

\subsection{Numerical evaluation of the current vertex}

\begin{figure}[ht!]
    \centering
    \includegraphics[width=\columnwidth]{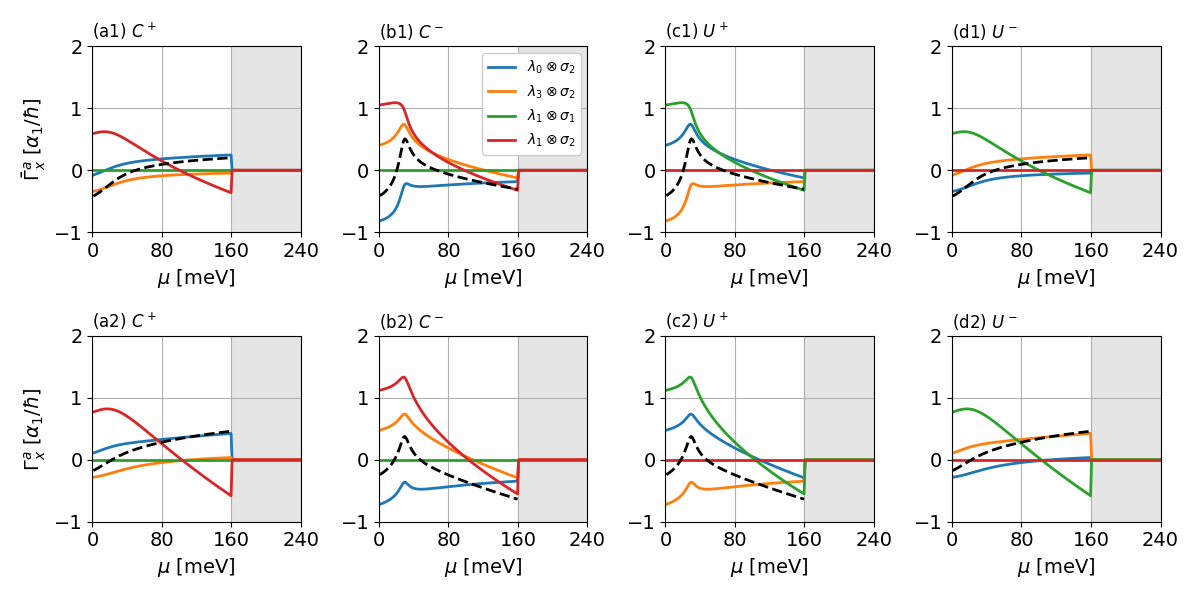}
    \caption{Components of the effective bare ${\bar \Gamma}_{x,a}$ (a1-d1), and dressed $\Gamma_{x,a}$ (a2-d2), charge current vertex in units of $\alpha_1/\hbar$, which represents the typical scale for $J_{soc,x}$. For the conventional case the only non zero components are $a=2$ ($\lambda_0\otimes\sigma_2$),  $a=14$ ($\lambda_3\otimes\sigma_2$), $a=6$ ($\lambda_1\otimes\sigma_2$). In the unconventional case instead they are $a=2$ ($\lambda_0\otimes\sigma_2$),  $a=14$ ($\lambda_3\otimes\sigma_2$), $a=5$ ($\lambda_1\otimes\sigma_1$). Dashed black lines refer to the sum of components $a=2$ ($\lambda_0\otimes\sigma_2$) and $a=14$ ($\lambda_3\otimes\sigma_2$) that constitute the single subband block when only the lower subbands are occupied ($\mu < 160$~meV).}
    \label{fig:effective_bare_vertex_x}
\end{figure}

The numerical evaluation of the charge current vertex begins with the evaluation of the effective bare vertex $\bar{\Gamma}_x$, which is the sum of the bare vertex $J_{soc,x}$ \eqref{zeta_decomposition_JSOC} and of the diagram with a single impurity line corresponding to the first iteration and whose integral has the matrix structure reported in Eq.~\eqref{integral_single_impurity_vertex}. In Fig.~\ref{fig:effective_bare_vertex_x} we show the non zero components as function of the Fermi energy. As in the case of the self-energy, two distinct regimes appear depending whether one or two bands are occupied. 
    In the first case, the finite components are the ones that appear in the $J_{soc,x}$ decomposition, Eq.~\eqref{zeta_decomposition_JSOC}, and are coupled by symmetry.

On the other hand when both bands are occupied, the effective bare vertex vanishes, as it happens in the single-band case\cite{raimondi:2001,raimondi:2005}. The vanishing of the effective bare vertex implies also the vanishing of the dressed vertex and the full current vertex reduces only to the first term in the right hand side of Eq.~\eqref{ansatz_vertex}.

In the regime when only one band is occupied, the effective bare vertex shows a remarkable non-monotonous behavior when bands coupling is allowed, both for the conventional and unconventional cases. This non-monotonous behavior could be expected in connection with the one of the density of states, which has a peak in the lowest band, when the dispersion flattens a bit around 50 meV.
The black dashed line in Fig.~\ref{fig:effective_bare_vertex_x} refers to the  sum of components $a=2$ ($\lambda_0\otimes\sigma_2$) and $a=14$ ($\lambda_3\otimes\sigma_2$), which describes the spin polarization along the y-axis in the single-subband block.

As a final general remark on the importance of vertex corrections one may state the following. When both bands are occupied, the vertex corrections have a dramatic impact in producing the vanishing of the matrix structure of the current vertex, no matter how weak the disorder scattering may be. This phenomenon is completely similar to what happens in the single-band case and remains true for both conventional and unconventional models as a result of the Rashba SOC. On the other hand, when only one band is occupied, the relevance of the vertex corrections mainly depends on the strength of the disorder scattering. 
The most important correction is due to the diagram with a single impurity line, which yields the effective bare vertex ${\bar \Gamma}_x$.
    In Figs.~\ref{fig:effective_bare_vertex_x}(a1-d1) we plot ${\bar \Gamma}_x$ in units of $\alpha_1/\hbar$, which is a typical scale for $J_{soc,x}$, which clearly shows that the vertex corrections typically leads to $|{\bar \Gamma}_x| < |J_{soc,x}|$. Interestingly, the vertex corrections might even flip the sign of the effective vertex with respect to $J_{soc,x}$ as a function of the Fermi energy.
Because we are in the regime of weak scattering the infinite resummation of ladder diagrams leading to the full vertex $\Gamma_x$ does not have a big impact on its numerical value, thus $\Gamma_x \approx \bar{\Gamma}_x$, as shown in Fig.~\ref{fig:effective_bare_vertex_x}(a2-d2).

\section{Charge-spin conversion}
\label{sec:numerics}
The current-induced spin polarization (CISP) is given by the response function $\chi_{yx}$ defined in Eq.~\eqref{kubo_formula}. In terms of the $\zeta_a$ matrices we have 
$S_y^0 = \zeta_2$.
Then by recalling the structure of the charge current vertex Eq.~\eqref{ansatz_vertex}, the response function reduces to two contributions
\begin{equation}
\chi_{yx}=\chi_{yx}^{(1)}+\chi_{yx}^{(2)},
\label{cisp_terms}
\end{equation}
where the first contribution is the simple bubble diagram with the spin vertex and the spin-independent velocity 
\begin{equation}
    \chi_{yx}^{(1)}=\frac{\hbar}{2\pi} \frac{(-e)\hbar}{2} \int \frac{d^2 k}{(2\pi)^2}
    {\rm Tr}\left[\zeta_2 G^R (\omega, {\bf k}) \frac{\hbar k_x}{m}  G^R (\omega, {\bf k}) \right],
    \label{cisp_1}
\end{equation}
and the second contribution originates from the spin-dependent part of the charge current vertex
\begin{equation}
    \chi_{yx}^{(2)}=
    \frac{\hbar}{2\pi} \frac{(-e)\hbar}{2} 
    \sum_b \int \frac{d^2 k}{(2\pi)^2}\Gamma_{x,b}
    {\rm Tr}\left[\zeta_2 G^R (\omega, {\bf k}) \zeta_b  G^R (\omega, {\bf k}) \right].
    \label{cisp_2}
\end{equation}
Both integrals in Eqs.~(\ref{cisp_1}-\ref{cisp_2}) can be evaluated with the technique previously developed and actually reduced to integrals already encountered in the derivation of the vertex corrections. The integral appearing in the first 
term $\chi_{yx}^{(1)}$ has been already found in Eq.~\eqref{bare_effective_vertex_x} for the effective bare vertex ${\bar\Gamma}_x$. On the other hand the integral appearing in the second term $\chi_{yx}^{(2)}$ can be expressed in terms of the matrix elements of the $L$-matrix, which carries the algebraic vertex corrections in Eq.~\eqref{linear_vertex_system}. As a result we get
\begin{align}
    \chi_{yx} &= -\frac{e\hbar}{\pi u_o^2}\left( {\bar \Gamma}_{x,2}-J_{SOC,x,2}\right)-\frac{e\hbar }{\pi u_0^2}
    \sum_b L_{2b} \Gamma_{x,b}=-\frac{e\hbar}{\pi u_o^2} \left( \Gamma_{x,2}-J_{SOC,x,2}\right).
    \label{cisp_total}
\end{align}
This is one of the main results of this paper and shows a remarkably compact expression, where only appear the dressed and bare vertices. 
Notice that to have the expression without the vertex corrections it is enough to make the replacement $\Gamma_x\rightarrow J_{SOC,x}$ before the last equality, where the relation from Eq.~ \eqref{linear_vertex_system} between the dressed $\Gamma_x$ and effective bare ${\bar\Gamma}_x$ has been used. 
To appreciate the importance of the vertex corrections, consider that, in the single-sub-band case, the dressed vertex vanishes exactly and hence the second term $\chi_{yx}^{(2)}$ is only different from zero when there are no vertex corrections. This implies that in the absence of vertex corrections one obtains a result which is smaller by a factor of two. In the present case, the second term of Eq.~\eqref{cisp_terms} may differ from zero even in the presence of vertex corrections, but only when one subband is occupied.

\begin{figure}
    \centering
    \includegraphics[width=\columnwidth]{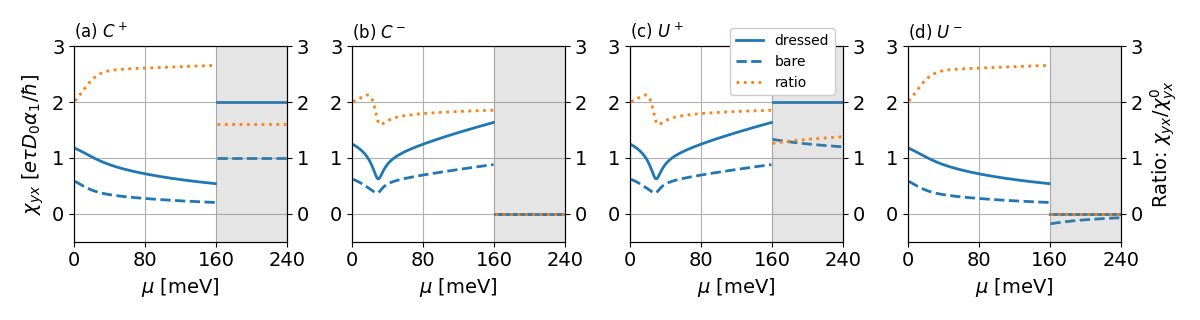}
    \caption{
    Current-induced spin polarization with ($\chi_{yx}$) and without ($\chi_{yx}^0$) vertex corrections, in units of its single subband limit, as function of the Fermi energy. In all cases we find $\chi_{yx} \approx 2\chi_{yx}^0$ due to the vertex corrections. 
    }
    \label{fig:CISP}
\end{figure}

The numerical evaluation of the CISP is shown in Fig.~\ref{fig:CISP} and it shows that the vertex corrections are always important yielding a factor around 2 between the dressed and bare vertex case. 
In the regime when both subbands are occupied, the dressed vertex vanishes (cf.~Fig.~\ref{fig:effective_bare_vertex_x}) and then only the first term in Eq.~\eqref{cisp_total} contributes. Furthermore, by considering 
the last equality in Eq.~\eqref{cisp_total}, the CISP is expressed only in terms of the bare vertex component $J_{SOC,x,2}$, which vanishes (see Eq.~\eqref{bare_charge_current_x}) when $\alpha_1 =-\alpha_2$ as it is evident in panels (b) and (d) in Fig.~\ref{fig:CISP}.
In this regime, for general values of the intra-band Rashba SOC one obtains the simple result
\begin{equation}
  \chi_{yx}=  -\frac{e\hbar}{\pi u_o^2} \left( -J_{SOC,x,2}\right)=e\tau D_0 \frac{\alpha_1+\alpha_2}{\hbar},
  \label{two_band_Edelstein}
\end{equation}
where the we have used the expression $\tau^{-1}=2\pi D_0 u_0^2 /\hbar$ for the scattering time, and the explicit expression from Eq.~\eqref{bare_charge_current_x} of the bare vertex component $J_{SOC,x,2}$. Hence, Eq.~\eqref{two_band_Edelstein} generalizes to two bands the well known Edelstein result\cite{edelstein:1990} with  exactly a factor of 2 between the dressed and bare vertex case. We stress that this result  is crucially obtained by considering the occupations of both bands.
At lower Fermi energies, when only the first sub-band is occupied, the second term $\chi_{yx}^{(2)}$ contributes and the CISP is  smaller and ratio between the dressed and bare vertex case varies around 2. As predicted by the last equality in Eq.~\eqref{cisp_total}, the behavior of the CISP is controlled by the dressed vertex $\Gamma_{x,2}$. In particular, the non-monotonous behavior observed in the vertex of Fig.~\ref{fig:effective_bare_vertex_x} is clearly reproduced in Fig.~\ref{fig:CISP}, for panels (b) and (c). Such non-monotonous behavior can be traced back,as already remarked, to the one observed in density of states in Fig.~\ref{fig:BandSpin}, where in columns two and three there is a flattening of the dispersion.

\begin{figure}
    \centering
    \includegraphics[width=\columnwidth]{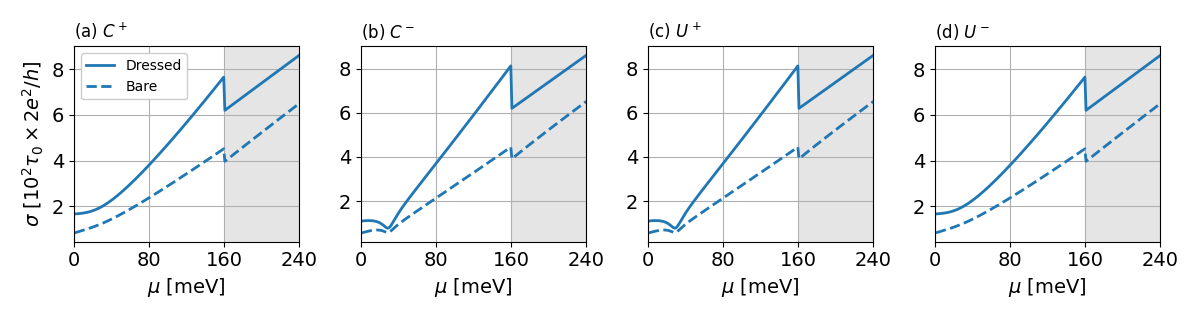}
    \caption{
    Dressed (solid) and bare conductivity $\sigma$ as a function of the Fermi energy.
    }
    \label{fig:ChargeCurrent}
\end{figure}

\begin{figure}
    \centering
    \includegraphics[width=\columnwidth]{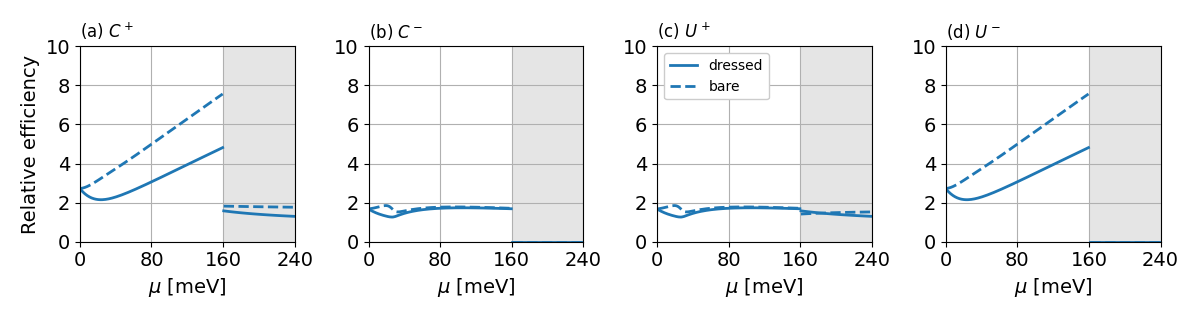}
    \caption{
    Dressed and bare IEE relative efficiencies $\sigma/\chi_{yx}$ normalized by the ratio in the uncoupled limit ($\eta = 0$). Both dressed and bare efficiencies are normalized by the dressed ratio with $\eta=0$.
    }
    \label{fig:LambdaIEE}
\end{figure}


In order to evaluate the charge-to-spin conversion efficiency, we need to evaluate the longitudinal electrical conductivity by the charge current-charge current response function
\begin{equation}
\sigma =  \frac{\hbar e^2}{2\pi}\int \frac{d^2 k}{(2\pi)^2}
{\rm Tr}\left[ J_x^0  ({\bf k}) G^R(\omega, {\bf k}) J_x ({\bf k})G^A(\omega, {\bf k})\right],
    \label{conductivity}
\end{equation}
where $J_x^0({\bf k})$ and 
$J_x({\bf k})$ are given by Eqs.~\eqref{bare_charge_current_x} and \eqref{ansatz_vertex}, respectively. By evaluating the integrals and using again the relation between the dressed, bare and effective bare vertex (cf. Eqs.~(\ref{bare_effective_vertex_x}, \ref{linear_vertex_system})), one finds 
\begin{align}
    \sigma &=
    e^2\sum_{sn}\frac{\hbar^2k^2_{sn}(\mu) D_{sn} }{4m^2\Gamma_{sn}/\hbar}
    +
    4e^2\tau D_0 
    \left[
        \sum_a 
        (\bar{\Gamma}_{x,a}-J_{SOC,x,a})
        (     \Gamma_{x,a} +J_{SOC,x,a})
        +
        \sum_{ab}
        J_{SOC,x,a} L_{ab} \Gamma_{x,a}
    \right],
\\
    \sigma &= e^2\sum_{sn}\frac{\hbar^2k^2_{sn}(\mu) D_{sn} }{4m^2\Gamma_{sn}/\hbar}
    +
    4e^2\tau D_0
    \sum_{a}
    \left( \Gamma_{x,a}{\bar\Gamma}_{x,a}-J_{SOC, x, a}J_{SOC, x, b}\right).
    \label{conductivity_final}
\end{align}
Above, the first expression is general, while to get the second expression we use the relation from Eq.~\eqref{linear_vertex_system}, which is valid only when we account for vertex corrections. To obtain the conductivity without vertex corrections, one simply has to replace $\Gamma_{x,a} \rightarrow J_{SOC,x,a}$ in the first expression for $\sigma$ above.
In both cases, the first term is the usual Drude expression summed over all the bands, as it can be seen by identifying $(\hbar^2k^2_{sn}(\mu) /m^2) D_{sn}/(4\Gamma_{sn}/\hbar )$ as the diffusion coefficient of band $sn$. The second term in Eq.~\eqref{conductivity_final} shows again a remarkable simplicity when expressed in terms of the dressed, bare and effective bare vertices. From this expression is also clear that this second term is at least quadratic in the SOC. The numerical results for $\sigma$ are shown in Fig.~\ref{fig:ChargeCurrent}, which clearly shows significant effects from the vertex corrections.

While Eqs.~(\ref{cisp_total}, \ref{conductivity_final}) provide clearly the spin polarization response and the electrical current response to an applied d.c. electric field, the issue of the charge-to-spin conversion efficiency is much less obvious. In the pioneering experiment by Sanchez et al. \cite{sanchez:2013}, where a spin current was pumped from  a  ferromagnetic metal into a bismuth-silver bi-layer, the dimensional efficiency $\lambda_{IEE}=j_c^{2D}/j_s^{3D}$ was introduced as the ratio of the induced two-dimensional interface charge current and the three-dimensional spin current flowing perpendicular to the interface of the bi-layer. We finally notice that, for the sake of convenience,  the spin current is defined as carrying a unit charge $e$ instead of carrying units of $\hbar/2$. 

In the work by Song et al. \cite{Song2021}, a different dimensionful definition of conversion efficiency $\lambda_{IEE}=j_c/j_s$ is adopted, where both the charge and spin current are two-dimensional quantities  and $j_s= e\langle S_y\rangle /\tau_F$. Also in this case the efficiency has the dimension of length due to the extra velocity factor in the definition of the charge current and the inverse momentum relaxation time factor in the spin current. In our opinion the choice of what conversion efficiency $\lambda_{IEE}$ to adopt cannot be decided in general, but must be selected depending on what experimental setting is under consideration. In our Fig.~\ref{fig:LambdaIEE} we have adopted as a measure of the efficiency the ratio $\sigma/\chi_{yx}$ normalized by the ratio to the uncoupled-band limit ($\eta =0$).

\begin{figure}[ht!]
    \centering
\includegraphics[width=\columnwidth]{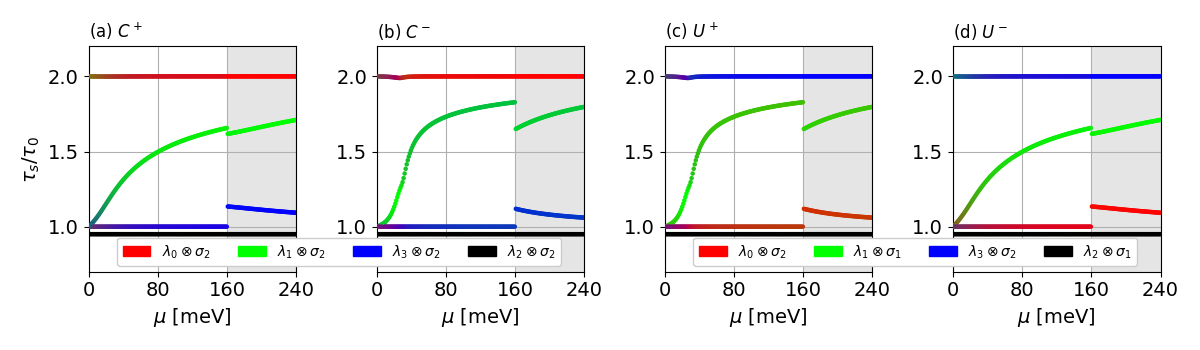}
    \caption{The relaxation times for the various observables in the block associated to spin polarization. In red the relaxation rate for the total spin polarization $\zeta_2 =\lambda_0\otimes \sigma_2$. 
    In all panels, the black line has $\tau_S/\tau_0 = 1$, but it was shifted downwards for clarity.
    }
    \label{fig:spin_lifetimes}
\end{figure}

To conclude this section, we report in Fig.~\ref{fig:spin_lifetimes} the spin relaxation times. These are provided by the eigenvalues of the $({\bf 1}-L)$-matrix of the algebraic vertex equation \eqref{linear_vertex_system}. To appreciate this, consider the vertex equation at finite external frequency $\Omega$, as due, for instance, to an a.c. electromagnetic field. The retarded and advanced Green's function acquire different frequencies, $\omega \rightarrow \omega +\Omega /2$ and $\omega \rightarrow \omega -\Omega /2$, respectively. When performing the momentum integral defining the matrix elements of the $L$-matrix, (cf. Eq.~\eqref{L-matrix}) one obtains a term proportional to $\Omega \tau_0 \ll 1$, which would give rise, after Fourier antitransforming back to real times, to a time derivative. Hence the eigenvalues of the $({\bf1 }-L)$-matrix would yield the relaxation rates in units of the inverse scattering time $\tau_0.$ 

In Fig.~\ref{fig:spin_lifetimes} we show the eigenvalues of the $({\bf 1}-L)^{-1}$ relative to the block to whom the spin polarization $\zeta_2 =\lambda_0 \otimes \sigma_2$ belongs. 
The figure uses a color code such that each observable has a finite weight in each eigenvalue. For instance in the panel (a) of Fig.~\ref{fig:spin_lifetimes} there is an eigenvalue with a lifetime around 2, which mostly coincides with the spin polarization $\lambda_0 \otimes \sigma_2$.
Such a value of 2 can be understood by recalling the single-band case. In this latter case, in the presence of the Rashba SOC, the spin relaxation time, $\tau_s$ for the Edelstein polarization is the well known Dyakonov-Perel spin relaxation time in the diffusive regime\cite{raimondi:2006}. In the deep diffusive regime, when the disorder broadening is larger than the spin splitting due to the SOC, the spin relaxation time is much longer than the quasiparticle relaxation time $\tau_0$, i.e. $\tau_s \gg \tau_0$. 
This is due to the fact that several impurity scattering events are necessary to relax the initial spin orientation.  For weak scattering, as it is considered here, spin precession may occur in between two impurity scattering events, and $\tau_s$ becomes of the order of the relaxation time $\tau_0$. In this, almost Elliot-Yafet regime, the single-band model predicts the exact relation $\tau_s=2 \tau_0$\cite{raimondi:2006}. 
In the conventional model (see panels (a) and (b) of Fig.~\ref{fig:spin_lifetimes}) a spin mode lifetime, consisting mostly of $\lambda_0 \otimes \sigma_2$  with the value of 2 in units of $\tau_0$ is always present, signalling that the spin dynamics of the two-band model behaves similarly to the single-band case. This is not surprising considering the spin texture for this case as shown in Fig.~\ref{fig:BandSpin} in the corresponding panels.

In the unconventional model, on the other hand, there is a again a spin mode with lifetime of 2, but its composition is markedly different, consisting mostly of an interband spin mode $\lambda_3 \otimes \sigma_2$ with spin polarization along the y-axis, but staggered between subbands.
This is an evident signal of a completely different spin dynamics, a result of the spin admixture of the energy bands.

\section{Conclusions}
\label{sec:summary}

In this paper we have studied the CISP and the charge-to-spin conversion efficiency in the two-band model in the presence of both intra- and interband Rashba SOC.  We have considered both conventional and unconventional interband Rashba coupling with the aim to analyze whether non-trivial spin texture may produce a more efficient charge-to-spin conversion. The CISP is a non-equilibrium phenomenon which is affected by the impurity disorder scattering relying on the deformation of the Fermi circle in the presence on an applied electric field. For this reason we have taken into account disorder scattering by means of the standard impurity diagrammatic technique. The evaluation of the rainbow diagram for the self-energy and the ladder diagram for the vertex has been carried out by fully exploiting the symmetry properties of the model, which allow to reduce the general algebraic vertex equation from dimension 16 by 16 to 4 by 4 blocks.  We have found that generically vertex corrections, at the level of the Born approximation, are important when both bands are occupied, whereas they are typically a small correction at lower Fermi energies when only one band is occupied.

We have found that the conventional or unconventional character of the inter-band SOC plays a role together with the relative signs of the intra-band SOC in the two subbands. 
Furthermore, for both types of inter-band SOC, it is important to take into account the occupation of all the bands for a consistent treatment of disorder scattering.
Hence, from the point of view of the efficiency, the unconventional model does not show itself better that the conventional one. On the other hand, it is also true that the unconventional model hosts a far richer spin dynamics that cannot be reduced to the one of the single-band case. This is a marked difference among the two models. We finally conclude by pointing out possible developments of the present analysis.
One direction is to consider stronger disorder, which is expected to require a fully self-consistent Born approximation, which can be carried out along the lines shown here, but is expected to be computationally more demanding. A second interesting direction is to expand the investigation of the complex spin dynamics of the unconventional model. We leave these promising paths to future work.

\appendix
\section{Derivation of the models}
\label{app:models}

The conventional case refers to a model that can be derived from two-subbands GaAs quantum wells grown along the [001] zincblende direction~\cite{calsaverini:2008}. If the quantum well is symmetric, its crystal lattice is invariant under the $D_{2D}$ point group. But in general, for an asymmetric well or in the presence of external fields, the structural inversion asymmetry (SIA) yields the $C_{2V} = \{C_{2}(z),M_{y}\}$ point group. Here $C_{2}(z)$ and $M_{y}$ are, respectively, the group generators referring to a $\pi$ rotation around the $z$ axis, and the mirror $y\rightarrow -y$, with $x=[110]$ and $y=[1\bar{1}0]$. In this case, both subband envelope functions transform as the $A_{1}(S)\oplus A_{1}(Z)$ irreps, and including spin we get $2A_{1}\otimes D_{1/2}=2\Gamma_{5}$, where $\Gamma_{5}=D_{1/2}$ is the pure spinor irrep. Therefore, we can label the basis set $\ket{j,\sigma}$ and order it as $\{\ket{1\uparrow},\ket{1\downarrow},\ket{2\uparrow},\ket{2\downarrow}\}$. Here $j=\{1,2\}$ labels the subbands, and $\sigma=\{\uparrow,\downarrow\}$ the spin. From these states, we obtain the representations for the group generators $C_{2}(z)=\lambda_{0}\otimes R_{2}(z)$, $M_{y}=-\lambda_{0}\otimes R_{2}(y)$, and for time-reversal symmetry $T=-i\lambda_{0}\otimes\sigma_{y}K$. Here $\lambda_{0}, \lambda_{x}, \lambda_{y}, \lambda_{z}$ are the identity and Pauli matrices in the subband space and $R_{n}(\hat{u})={\rm exp}(i\frac{\pi}{n}\hat{u}\cdot\bm{\sigma})$ is the spin rotation by $2\pi/n$ over the unit vector $\hat{u}$. From these representations, using the method of invariants via the \texttt{Qsymm} code \cite{Varjas2018}, we obtain the conventional case Hamiltonian $H_C$ as
\begin{align}
    H_{C} =
    \begin{pmatrix}
    \varepsilon_{1} & -i\alpha_{1}k_{-} & 0 & -i\eta_C k_{-}
    \\
    i\alpha_{1}k_{+} & \varepsilon_{1} & i\eta_C k_{+} & 0
    \\
    0 & -i\eta_C k_{-} & \varepsilon_{2} & -i\alpha_{2}k_{-}
    \\
    i\eta_C k_{+} & 0 & i\alpha_{2}k_{+} & \varepsilon_{2}
\end{pmatrix}.
\end{align}
Here, for each subband $j$, $\varepsilon_j = \varepsilon_{j}^{0}+\frac{\hbar^2}{2m}k^2$, $\varepsilon_{j}^{0}$ are the band edges, the effective mass $m$ is assumed to be the same in both subands, $\alpha_j$ is the intra-subband Rashba SOC, $\eta_C$ is the inter-subband Rashba SOC, $\bm{k}=(k_x, k_y)$ is the in-plane quasi-momentum, and $k_\pm = k_x \pm ik_y$.

In contrast, the unconventional case occurs in 2D materials that transform as the $C_{3V}$ group at the $\Gamma$ point as, for instance, the monolayer OsBi$_2$ discussed in Ref.~\citeonline{Song2021}. There, the orbitals of the two relevant bands transform as the $E$ irrep of the $C_{3V}$ single group, and including spin it splits into $2E\otimes D_{1/2}=2(\Gamma_{4}\oplus\Gamma_{5}\oplus\Gamma_{6})$, where $\Gamma_{4}=D_{1/2}$ is the pure spinor irrep, and $\Gamma_{5}\oplus\Gamma_{6}$ are 1D irreps that form Kramer's pairs under time-reversal symmetry. More specifically, the single group orbital representations $E^j$ for each subband $j=\{1,2\}$ are $E^1 = \ket{X_\pm Z}$ and $E^2 = \ket{XY\pm \frac{i}{2}(X^2-Y^2)}$, where $X_\pm = X\pm iY$. Including spin, the $E^1$ orbitals splits into $\Gamma^1_5\oplus\Gamma^1_6 = \{\ket{X_+Z\uparrow}, \ket{X_-Z\downarrow}\}$, and $\Gamma_4^1 = \{\ket{X_-Z\uparrow}, \ket{X_+Z\downarrow}\}$. For the other subband, $E^2$ splits into $\Gamma^2_5\oplus\Gamma^2_6 = \{\ket{XY+\frac{i}{2}(X^2-Y^2)\uparrow}, \ket{XY-\frac{i}{2}(X^2-Y^2)\downarrow}\}$, and $\Gamma_4^2 = \{\ket{XY-\frac{i}{2}(X^2-Y^2)\uparrow}, \ket{XY+\frac{i}{2}(X^2-Y^2)\downarrow}\}$.

Interestingly, similarly to the conventional case, here the unconventional model also arises from the spinor irreps $\Gamma_4^1\oplus\Gamma_4^2$, but under different constraints from the $C_{3V}$ group. Sorting this basis set as $\Gamma_{4}^1\oplus\Gamma_{4}^2 = \{\ket{X_-Z\uparrow},\ket{X_+Z\downarrow},\ket{XY-\frac{i}{2}(X{{}^2}-Y{{}^2})\uparrow},\ket{XY+\frac{i}{2}(X{{}^2}-Y{{}^2})\downarrow}\} \equiv \{\ket{1\uparrow},\ket{1\downarrow},\ket{2\uparrow},\ket{2\downarrow}\}$, the group generators read as $C_3(z) = -\lambda_{0}\otimes R_{3}(z)$, $M_y = \lambda_{z}\otimes(i\sigma_{y})$, and the time-reversal operator is $T = \lambda_{0}\otimes(i\sigma_{y})K$. These lead to the unconventional Hamiltonian $H_U$, which read as
\begin{align}
    H_U = 
    \begin{pmatrix}
    \varepsilon_{1} & -i\alpha_{1}k_{-} & 0 & \eta_U k_{-}
    \\
    i\alpha_{1}k_{+} & \varepsilon_{1} & \eta_U k_{+} & 0
    \\
    0 & \eta_U k_{-} & \varepsilon_{2} & -i\alpha_{2}k_{-}
    \\
    \eta_U k_{+} & 0 & i\alpha_{2}k_{+} & \varepsilon_{2}
\end{pmatrix}.
\end{align}
Here $\eta_U$ is the intersubband Rashba SOC for the unconventional case, and the other quantities match the definitions from $H_C$ above.

\section{The ``bubble'' integrals}
\label{app:bubble_integrals}

When evaluting the vertex corrections and the response functions, one encounters integrals of the type (see Eq.~\eqref{L-matrix})
\begin{equation}
    L_{ab}= u_0^2 \int \frac{d^2 k}{(2\pi)^2} {\rm Tr} \left[ \zeta_a G^R(\omega, {\bf k}) \zeta_b G^A(\omega, {\bf k})\right].
    \label{appb:Lab_definition}
\end{equation}
By setting $g^{R(A)}_{sn}(\omega, k)=(\omega -E_{sn}(k))^{-1}$ and recalling the spectral decomposition shown in  Eq.~\eqref{selfenergy} one obtains
\begin{equation}
     L_{ab}= u_0^2\sum_{sn s' n'} \int \frac{d^2 k}{(2\pi)^2} g^{R}_{sn}(\omega, k) g^{A}_{s'n'}(\omega, k)
     {\rm Tr} \left[\zeta_a P_{sn}({\bf k}) \zeta_b P_{s'n'}({\bf k}) \right].
\end{equation}
We notice that all the angle dependence is within the projectors. We may then use the decomposition from Eq.~\eqref{angle_dependence_projector} and perform at once the integral over $\theta$
\begin{equation}
     P^{ab}_{sns'n'}(k) \equiv \int_0^{2\pi} \frac{d\theta}{2\pi}
    {\rm Tr} \left[\zeta_a P_{sn}({\bf k}) \zeta_b P_{s'n'}({\bf k}) \right]=
    {\rm Tr}\left[\zeta_a P^0_{sn}\zeta_b P^0_{s'n'} +\zeta_a P^+_{sn}\zeta_b P^-_{s'n'} +\zeta_a P^-_{sn}\zeta_b P^+_{s'n'}  \right],
    \label{appb:trace_lab}
\end{equation}
where we have omitted for brevity the dependence on the absolute value of the momentum $P^0_{sn}\equiv P^0_{sn} (k)$.

The matrix elements $L_{0a}=L_{a0}$ acquire a simpler expression because the closeness of the two Green's functions under the trace symbol  allows to exploit the property of the projectors and one gets
\begin{equation}
     L_{0a}= u_0^2\sum_{sn } \int \frac{d^2 k}{(2\pi)^2} g^{R}_{sn}(\omega, k) g^{A}_{sn}(\omega, k)
     {\rm Tr} \left[ P_{sn}({\bf k}) \zeta_a  \right]=u_0^2\sum_{sn }
     \int \frac{k dk}{2\pi}g^{R}_{sn}(\omega, k) g^{A}_{sn}(\omega, k) {\rm Tr} \left[ P^0_{sn}( k) \zeta_a  \right].
\end{equation}
The above matrix element can only differ from zero when the index $a$ belongs to the set of $\zeta$ matrices appearing in the decomposition of $P^0_{sn}(k)$ as shown in Eqs.~(\ref{angle_P0_integrated_projector_C}, \ref{angle_P0_integrated_projector_U}).

To evaluate the matrix $L_{ab}$ we further need the integral over the absolute value of the momentum
\begin{equation}
    {\cal G}^{(2)}_{sn s'n'}=\int \frac{k dk}{2\pi}g^{R}_{sn}(\omega, k) g^{A}_{s'n'}(\omega, k).
\end{equation}
The above integral depends on the comparison between the energy differences $E_{sn}-E_{s'n'}$ and the disorder-induced broadening $\Gamma_{sn}=-{\rm Im} \Sigma^R_{sn}$, $\Gamma_{s'n'}=-{\rm Im} \Sigma^R_{s'n'}$, $\Sigma^R_{sn}$ being 
the elements of the self-energy in the diagonal basis.
For the case $E_{sn}-E_{s'n'} \ll \Gamma_{sn} \sim \Gamma_{s'n'}$, one has
\begin{equation}
     {\cal G}^{(2)}_{sn s'n'}=\frac{\pi}{2}\left( \frac{D_{sn}}{\Gamma_{sn}}+ \frac{D_{s'n'}}{\Gamma_{s'n'}}\right).
     \label{g2degenerate}
\end{equation}
In the opposite case of large energy separation $E_{sn}-E_{s'n'} \gg \Gamma_{sn} \sim \Gamma_{s'n'}$ we have instead
\begin{equation}
     {\cal G}^{(2)}_{sn s'n'}=-i\pi\left( \frac{D_{sn}}{\Delta_{sn}}+ \frac{D_{s'n'}}{\Delta_{s'n'}}\right),
     \label{g2gapped}
\end{equation}
where $\Delta_{sn}=E_{sn}-E_{s'n'}$ with $E_{sn}\equiv E_{sn}(k_{sn}(\mu))$ and 
$E_{s'n'}\equiv E_{s'n'}(k_{sn}(\mu))$ and a similar expression for  $\Delta_{s'n'}$.

For the case in Eq.~\eqref{g2degenerate}, we see that ${\cal G}^{(2)}_{sn s'n'} \propto u_0^{-2}$, which exactly cancels the $u_0^2$ prefactor in $L_{ab}$ and gives its leading order contribution. On the other hand, for the case in Eq.~\eqref{g2gapped}, the $u_0^2$ term appears only in the imaginary part of $\Delta_{sn}$, which is assumed to be small, and the leading contribution of ${\cal G}^{(2)}_{sn s'n'}$ in this case is independent of $u_0^2$. Consequently, its contribution to the $L_{ab}$ matrix is of order $u_0^2$, which we neglect within the first Born approximation.

\acknowledgments        

 G.J.F. acknowledges support from the Brazilian funding agencies CNPq, CAPES and FAPEMIG (Grant PPM-00798-18).
J.Y.F. acknowledges support by the National Natural Science Foundation of China (Grants No. 12274256 and
No. 11874236) and the Major Basic Program of Natural Science Foundation of Shandong Province (Grant No.
ZR2021ZD01).
One of the authors (J.Y.F.) thanks Rui Song and Ning Hao for helpful discussions.

\bibliography{biblio} 

\begin{thebibliography}{10}

\bibitem{awschalom:2002}
Awschalom, D., Loss, D., and Samarth, N.,  [{\em Semiconductor Spintronics and
  Quantum Computation}{\nolinebreak\hspace{0.1em}]}, Springer, New York (2002).

\bibitem{zutic:2004}
\v{Z}uti\'c, I., Fabian, J., and Sarma, S.~D., ``Spintronics: Fundamentals and
  applications,'' {\em Rev. Mod. Phys.}~{\bf 76},  323 (2004).

\bibitem{bernevig2:2006}
Bernevig, B.~A., Hughes, T.~L., and Zhang, S.~C., ``Quantum spin {Hall} effect
  and topological phase transition in {HgTe} quantum wells,'' {\em
  Science}~{\bf 314},  1757 (2006).

\bibitem{lutchyn:2010}
Lutchyn, R.~M., Sau, J.~D., and Sarma, S.~D., ``Majorana fermions and a
  topological phase transition in semiconductor-superconductor
  heterostructures,'' {\em Phys. Rev. Lett.}~{\bf 105},  077001 (2010).

\bibitem{oreg:2010}
Oreg, Y.~Refael, G. and Oppen, F.~v., ``Helical liquids and majorana bound
  states in quantum wires,'' {\em Phys. Rev. Lett.}~{\bf 105},  177002 (2010).

\bibitem{geim:2013}
Geim, A.~K. and Grigorieva, I.~V., ``{Van der Waals heterostructures},'' {\em
  Nature}~{\bf 499} (2013).

\bibitem{gmitra:2015}
Gmitra, M. and Fabian, J., ``Graphene on transition-metal dichalcogenides: A
  platform for proximity spin-orbit physics and optospintronics,'' {\em Phys.
  Rev. B}~{\bf 92},  155403 (Oct 2015).

\bibitem{weng:2015}
Weng, H.~M., Fang, C., Fang, Z., Bernevig, B.~A., and Dai, X., ``{Weyl
  Semimetal Phase in Noncentrosymmetric Transition-Metal Monophosphides},''
  {\em Phys. Rev. X}~{\bf 105},  011029 (2015).

\bibitem{burkov:2015}
Burkov, A.~A., ``{Chiral anomaly and transport in Weyl metals},'' {\em J.
  Phys.: Condens. Matter}~{\bf 27},  113201 (feb 2015).

\bibitem{fu:2016}
Fu, J.~Y., Penteado, P.~H., Hachiya, M.~O., Loss, D., and Egues, J.~C.,
  ``Persistent skyrmion lattice of noninteracting electrons with spin-orbit
  coupling,'' {\em Phys. Rev. Lett.}~{\bf 117},  226401 (2016).

\bibitem{dettwiler:2017}
Dettwiler, F., Fu, J., Mack, S., Weigele, P.~J., Egues, J.~C., Awschalom,
  D.~D., and Zumb\"uhl, D.~M., ``Stretchable persistent spin helices in gaas
  quantum wells,'' {\em Phys. Rev. X}~{\bf 7},  031010 (Jul 2017).

\bibitem{zhao:2023}
Zhao, N., Duan, Y.~H., Yang, H., Li, X., Liu, W., Zhao, J.~H., Han, S.~X.,
  Shen, K., Hao, N., Fu, J.~Y., and Zhang, P., ``Two copies of persistent spin
  helices with stretching pitch and compensating helicity,'' {\em Phys. Rev.
  B}~{\bf 107},  205407 (2023).

\bibitem{edelstein:1990}
Edelstein, V.~M., ``Spin polarization of conduction electrons induced by
  electric current in two-dimensional asymmetric electron systems,'' {\em Solid
  State Commun.}~{\bf 73},  233 (1990).

\bibitem{aronov:1989}
Aronov, A.~G. and Lyanda-Geller, Y.~B., ``Nuclear electric resonance and
  orientation of carrier spin by an electric field,'' {\em JETP Lett.}~{\bf
  50},  431 (1989).

\bibitem{ganichev:2002}
Ganichev, S.~D., Ivchenko, E.~L., Belkov, V.~V., Tarasenko, S.~A., Sollinger,
  M., Weiss, D., Wegscheider, W., and Prettl, W., ``Spin-galvanic effect,''
  {\em Nature}~{\bf 417},  153 (2002).

\bibitem{ivchenko:1978}
{Ivchenko}, E.~L. and {Pikus}, G.~E., ``{New photogalvanic effect in gyrotropic
  crystals},'' {\em JETP Lett.}~{\bf 27},  604 (June 1978).

\bibitem{vorobev:1979}
{Vorob'ev}, L.~E., {Ivchenko}, E.~L., {Pikus}, G.~E., {Farbshte{\v i}n}, I.~I.,
  {Shalygin}, V.~A., and {Shturbin}, A.~V., ``{Optical activity in tellurium
  induced by a current},'' {\em JETP Lett.}~{\bf 29},  441 (Apr. 1979).

\bibitem{huang:2016}
Huang, C., Chong, Y.~D., and Cazalilla, M.~A., ``Direct coupling between charge
  current and spin polarization by extrinsic mechanisms in graphene,'' {\em
  Phys. Rev. B}~{\bf 94},  085414 (Aug 2016).

\bibitem{offidani:2017}
Offidani, M., Milletar\`{\i}, M., Raimondi, R., and Ferreira, A., ``Optimal
  charge-to-spin conversion in graphene on transition-metal dichalcogenides,''
  {\em Phys. Rev. Lett.}~{\bf 119},  196801 (Nov 2017).

\bibitem{lin:2019}
Lin, Y.-H., Huang, C., Offidani, M., Ferreira, A., and Cazalilla, M.~A.,
  ``Theory of spin injection in two-dimensional metals with proximity-induced
  spin-orbit coupling,'' {\em Phys. Rev. B}~{\bf 100},  245424 (Dec 2019).

\bibitem{ghiasi:2019}
Ghiasi, T.~S., Kaverzin, A.~A., Blah, P.~J., and van Wees, B.~J.,
  ``{Charge-to-Spin Conversion by the Rashba–Edelstein Effect in
  Two-Dimensional van der Waals Heterostructures up to Room Temperature},''
  {\em Nano Lett.}~{\bf 19}(9),  5959--5966 (2019).
\newblock PMID: 31408607.

\bibitem{benitez:2020}
Ben{\'\i}tez, L.~A., Savero~Torres, W., Sierra, J.~F., Timmermans, M., Garcia,
  J.~H., Roche, S., Costache, M.~V., and Valenzuela, S.~O., ``{Tunable
  room-temperature spin galvanic and spin Hall effects in van der Waals
  heterostructures},'' {\em Nat. Mater.}~{\bf 19}(2),  170--175 (2020).

\bibitem{monaco:2021}
Monaco, C., Ferreira, A., and Raimondi, R., ``{Spin Hall and inverse spin
  galvanic effects in graphene with strong interfacial spin-orbit coupling: A
  quasi-classical Green's function approach},'' {\em Phys. Rev. Res.}~{\bf 3},
  033137 (Aug 2021).

\bibitem{rashba:1984}
Bychkov, Y.~A. and Rashba, E.~I., ``Properties of a {2D} electron gas with
  lifted spectral degeneracy,'' {\em {JETP} Letters}~{\bf 39},  78 (1984).

\bibitem{dresselhaus:1955}
Dresselhaus, G., ``Spin-orbit coupling effects in zinc blende structures,''
  {\em Phys. Rev.}~{\bf 100},  580 (1955).

\bibitem{walser:2012_2}
Walser, M.~P., Siegenthaler, U., Lechner, V., Schuh, D., Ganichev, S.~D.,
  Wegscheider, W., and Salis, G., ``Dependence of the {D}resselhaus spin-orbit
  interaction on the quantum well width,'' {\em Phys. Rev. B}~{\bf 86},  195309
  (2012).

\bibitem{fu:2015}
Fu, J.~Y. and Egues, J.~C., ``Spin-orbit interaction in {GaAs} wells: From one
  to two subbands,'' {\em Phys. Rev. B}~{\bf 91},  075408 (2015).

\bibitem{nitta:1997}
Nitta, J., Akazaki, T., Takayanagi, H., and Enoki, T., ``Gate control of
  spin-orbit interaction in an inverted
  {In$_{0.53}$Ga$_{0.47}$As/In$_{0.52}$Al$_{0.48}$As},'' {\em Phys. Rev.
  Lett.}~{\bf 78},  1335 (1997).

\bibitem{studer:2009}
Studer, M., Salis, G., Ensslin, K., Driscoll, D.~C., and Gossard, A.~C.,
  ``Gate-controlled spin-orbit interaction in a parabolic {GaAs/AlGaAs} quantum
  well,'' {\em Phys. Rev. Lett.}~{\bf 103},  027201 (2009).

\bibitem{sasaki:2014}
Sasaki, A., Nonaka, S., Kunihashi, Y., Kohda, M., Bauernfeind, T., Dollinger,
  T., Richter, K., and Nitta, J., ``Direct determination of spin-orbit
  interaction coefficients and realization of the persistent spin helix
  symmetry,'' {\em Nat. Nanotechnol.}~{\bf 9},  1748 (2014).

\bibitem{chuang:2015}
Chuang, P., Ho, S., Smith, L.~W., Sfigakis, F., Pepper, M., Chen, C., Fan, J.,
  Griffiths, J.~P., Farrer, I., Beere, H.~E., Jones, G. A.~C., Ritchie, D.~A.,
  and Chen, T., ``All-electric all-semiconductor spin field-effect
  transistors,'' {\em Nat. Nanotechnol.}~{\bf 10},  35 (2015).

\bibitem{koo:2009}
Koo, H.~C., Kwon, J.~H., Eom, J., Chang, J., Han, S.~H., and Johnson, M.,
  ``Control of spin precession in a spin-injected field effect transistor,''
  {\em Science}~{\bf 325},  1515 (2009).

\bibitem{datta:1990}
Datta, S. and Das, B., ``Electronic analog of the electro-optic modulator,''
  {\em App. Phys. Lett.}~{\bf 56},  665 (1990).

\bibitem{sinova:2015}
Sinova, J., Valenzuela, S.~O., Wunderlich, J., Back, C., and Jungwirth, T.,
  ``Spin {Hall} effects,'' {\em Rev. Mod. Phys.}~{\bf 87},  1213 (2015).

\bibitem{wunderlich:2010}
Wunderlich, J., Park, B., Irvine, A.~C., Z$\rm\hat{a}$rbo, L.~P., Rozkotov\'a,
  E., Nemec, P., Nov\'ak, V., Sinova, J., and Jungwirth, T., ``Spin {Hall}
  effect transistor,'' {\em Science}~{\bf 30},  1801 (2010).

\bibitem{calsaverini:2008}
Calsaverini, R.~S., Bernardes, E., Egues, J.~C., and Loss, D.,
  ``Intersubband-induced spin-orbit interaction in quantum wells,'' {\em Phys.
  Rev. B}~{\bf 78},  155313 (2008).

\bibitem{fu:2020}
Fu, J.~Y., Penteado, P.~H., Candido, D.~R., Ferreira, G.~J., Pires, D.~P.,
  Bernardes, E., and Egues, J.~C., ``Spin-orbit coupling in wurtzite
  heterostructures,'' {\em Phys. Rev. B}~{\bf 101},  134416 (2020).

\bibitem{hao:2015}
Hao, Y.~F., ``Spin-orbit interaction in multiple quantum wells,'' {\em J. Appl.
  Phys.}~{\bf 117},  013911 (2015).

\bibitem{kunihashi:2016}
Kunihashi, Y., Sanada, H., Gotoh, H., Onomitsu, K., Kohda, M., Nitta, J., and
  Sogawa, T., ``Drift transport of helical spin coherence with tailored
  spin--orbit interactions,'' {\em Nat. Commun.}~{\bf 7}(1),  10722 (2016).

\bibitem{Ferreira2017RW}
Ferreira, G.~J., Hernandez, F. G.~G., Altmann, P., and Salis, G., ``Spin drift
  and diffusion in one- and two-subband helical systems,'' {\em Phys. Rev.
  B}~{\bf 95},  125119 (Mar 2017).

\bibitem{weigele:2020}
Weigele, P.~J., Marinescu, D.~C., Dettwiler, F., Fu, J., Mack, S., Egues,
  J.~C., Awschalom, D.~D., and Zumb\"uhl, D.~M., ``Symmetry breaking of the
  persistent spin helix in quantum transport,'' {\em Phys. Rev. B}~{\bf 101},
  035414 (Jan 2020).

\bibitem{deassis:2021}
de~Assis, I.~R., Raimondi, R., and Ferreira, G.~J., ``Spin drift-diffusion for
  two-subband quantum wells,'' {\em Phys. Rev. B}~{\bf 103},  165304 (Apr
  2021).

\bibitem{Sergio2023OrbitalEdelstein}
M., S.~L., Henk, J., Mertig, I., and Johansson, A., ``{Spin and orbital
  Edelstein effect in a bilayer system with Rashba interaction},'' (2023).

\bibitem{bernardes2006spin}
Bernardes, E., Schliemann, J., Egues, J.~C., and Loss, D., ``Spin orbit
  interaction and zitterbewegung in symmetric wells,'' {\em Phys. Status Solidi
  C}~{\bf 3}(12),  4330 (2006).

\bibitem{EsmerindoPRL2007}
Bernardes, E., Schliemann, J., Lee, M., Egues, J.~C., and Loss, D.,
  ``Spin-orbit interaction in symmetric wells with two subbands,'' {\em Phys.
  Rev. Lett.}~{\bf 99},  076603 (2007).

\bibitem{Hernandez2016CISP}
Hernandez, F. G.~G., Ullah, S., Ferreira, G.~J., Kawahala, N.~M., Gusev, G.~M.,
  and Bakarov, A.~K., ``Macroscopic transverse drift of long current-induced
  spin coherence in two-dimensional electron gases,'' {\em Phys. Rev. B}~{\bf
  94},  045305 (Jul 2016).

\bibitem{Luengo2017Gate}
Luengo-Kovac, M., Moraes, F. C.~D., Ferreira, G.~J., Ribeiro, A. S.~L., Gusev,
  G.~M., Bakarov, A.~K., Sih, V., and Hernandez, F. G.~G., ``Gate control of
  the spin mobility through the modification of the spin-orbit interaction in
  two-dimensional systems,'' {\em Phys. Rev. B}~{\bf 95},  245315 (Jun 2017).

\bibitem{Hernandez2020Anisotropy}
Hernandez, F. G.~G., Ferreira, G.~J., Luengo-Kovac, M., Sih, V., Kawahala,
  N.~M., Gusev, G.~M., and Bakarov, A.~K., ``Electrical control of spin
  relaxation anisotropy during drift transport in a two-dimensional electron
  gas,'' {\em Phys. Rev. B}~{\bf 102},  125305 (Sep 2020).

\bibitem{Song2021}
Song, R., Hao, N., and Zhang, P., ``{Giant inverse Rashba-Edelstein effect:
  Application to monolayer ${\mathrm{OsBi}}_{2}$},'' {\em Phys. Rev. B}~{\bf
  104},  115433 (Sep 2021).

\bibitem{bentmann:2012}
Bentmann, H., Abdelouahed, S., Mulazzi, M., Henk, J., and Reinert, F., ``Direct
  observation of interband spin-orbit coupling in a two-dimensional electron
  system,'' {\em Phys. Rev. Lett.}~{\bf 108},  196801 (2012).

\bibitem{noguchi:2017}
Noguchi, R.~Kuroda, K., Yaji, K., Kobayashi, K., Sakano, M., Harasawa, A.,
  Kondo, T., Komori, F., and Shin, S., ``Direct mapping of spin and orbital
  entangled wave functions under interband spin-orbit coupling of giant
  {Rashba} spin-split surface states,'' {\em Phys. Rev. B}~{\bf 95},  041111(R)
  (2017).

\bibitem{shen:2014}
Shen, K., Vignale, G., and Raimondi, R., ``{Microscopic Theory of the Inverse
  Edelstein Effect},'' {\em Phys. Rev. Lett.}~{\bf 112},  096601 (Mar 2014).

\bibitem{raimondi:2001}
Raimondi, R., Leadbeater, M., Schwab, P., Caroti, E., and Castellani, C.,
  ``Spin-orbit induced anisotropy in the magnetoconductance of two-dimensional
  metals,'' {\em Phys. Rev. B}~{\bf 64},  235110 (Nov 2001).

\bibitem{schwab:2002}
{Schwab}, P. and {Raimondi}, R., ``{Magnetoconductance of a two-dimensional
  metal in the presence of spin-orbit coupling},'' {\em Eur. Phys. J. B}~{\bf
  25},  483--495 (2002).

\bibitem{raimondi:2005}
Raimondi, R. and Schwab, P., ``{Spin-Hall effect in a disordered 2D electron
  system},'' {\em Phys. Rev. B}~{\bf 71},  033311 (2005).

\bibitem{sanchez:2013}
S\'anchez, J. C.~R., Vila, L., Desfonds, G., Gambarelli, S., Attan\'e, J.~P.,
  Teresa, J. M.~D., Mag\'en, C., and Fert, A., ``{Spin-to-charge conversion
  using Rashba coupling at the interface between non-magnetic materials},''
  {\em Nature Commun.}~{\bf 4},  2944 (2013).

\bibitem{raimondi:2006}
Raimondi, R., Gorini, C., Schwab, P., and Dzierzawa, M., ``{Quasiclassical
  approach to the spin Hall effect in the two-dimensional electron gas},'' {\em
  Phys. Rev. B}~{\bf 74},  035340 (Jul 2006).

\bibitem{Varjas2018}
Varjas, D., Rosdahl, T.~O., and Akhmerov, A.~R., ``{Qsymm: algorithmic symmetry
  finding and symmetric Hamiltonian generation},'' {\em New J. Phys.}~{\bf 20},
   093026 (Sept. 2018).

\end{thebibliography}
\bibliographystyle{spiebib} 

\end{document}